\documentclass{pas}
\usepackage{preamble}
\usepackage{comment}

\newcommand{\objnamelong}{J085337.26+014303.6}
\newcommand{\objname}{J0853}
\newcommand{\ergscma}{erg\,s$^{-1}$\,cm$^{-2}$\,\AA$^{-1}$}
\newcommand{\ergscm}{erg\,s$^{-1}$\,cm$^{-2}$}
\newcommand{\ergs}{erg\,s$^{-1}$}
\def \ppxf {{\small P}PXF}

\def\kmsmpc {km$\,$s$^{-1}$Mpc$^{-1}$}
\def \kms {km$\,$s$^{-1}$}
\def \oiii {[O{\small~III}]}

\def \sii {[S{\small~II}]}
\def \nii {[N{\small~II}]}
\def \ha {H$\alpha$}
\def \hb {H$\beta$}
\def \msun {M$_\odot$}
\def \mbh {M_{\text{BH}}}
\def \dnfour {D$_n$(4000)}

\begin{document}

\lefttitle{Publications of the Astronomical Society of Australia}
\righttitle{S. M. Croom et al.}

\jnlPage{1}{4}
\jnlDoiYr{2026}
\doival{10.1017/pasa.xxxx.xx}

\articletitt{Research Paper}

\title{A recurring changing state AGN discovered in the SAMI Galaxy Survey and SDSS}

\author{Scott M. Croom$^1$, Rebecca McElroy$^{2,1,4}$, Dougal Dobie$^{1,3}$, Thomas Dunn$^1$, Yasmeen Janschek$^1$, Tyler Klinger$^1$, Shaun Parasher$^1$, Mitchell Spanner$^1$, Oliver J. Thorp$^1$}
\affil{$^1$Sydney Institute for Astronomy (SIfA), School of Physics, The University of Sydney, NSW 2006, Australia\\
$^2$Centre for Astrophysics, University of Southern Queensland, Toowoomba, QLD 4350, Australia\\
$^3$ARC Centre of Excellence for Gravitational Wave Discovery (OzGrav), Hawthorn, Victoria, 3122, Australia\\
$^4$Astrophysics and Space Technologies Research Centre, 12 Wally's Walk, Macquarie University, 2109, Australia
}

\corresp{S. M. Croom, Email: scott.croom@sydney.edu.au}

\citeauth{Croom S.M. et al., A recurring changing state AGN discovered in the SAMI Galaxy Survey and SDSS.  {\it Publications of the Astronomical Society of Australia} {\bf 00}, 1--12. https://doi.org/10.1017/pasa.xxxx.xx}

\history{(Received xx xx xxxx; revised xx xx xxxx; accepted 21 10 2026)}

\begin{abstract}
We present a detailed analysis of the AGN in galaxy \objnamelong\ at $z=0.0582$.  A combination of SDSS and SAMI Galaxy Survey data shows that this object has gone through at least one full changing-look AGN cycle, from type 1.9 in 2001 to type 1 in 2015 and 2021, and then type 1.9 again in 2022.  While the broad lines in this object show evidence of dust obscuration, detailed spectral analysis plus WISE photometry demonstrate that the changes in this object are related to a physical state change, not changing obscuration.  In fact, the WISE light curve suggests that this object may have briefly returned to a type 1.9 state in 2020.  In the optical, the host galaxy dominates the light (e.g.\ only $\sim 10$ per cent contribution from the AGN at 4000\AA\ within a central aperture), even in the high state.  In contrast, the brightest WISE W2-band measurements place the AGN at a similar luminosity to the entire host galaxy.  The broad Balmer lines show evidence of multiple components, with a weak very-broad component (FWHM $\simeq5000-6000$\,\kms) visible in the low type 1.9 state, and a stronger broad component of FWHM $\sim2000$\kms dominating in the high type 1 state.  We propose that this is caused by two spatially distinct regions within the broad line region.
\end{abstract}


\maketitle

\section{Introduction}

Active Galactic Nuclei (AGN) are the bright cores of galaxies powered by the infall of gas and dust onto a central supermassive black hole (SMBH) \citep{pringle_accretion_1973}. Traditionally, AGN are classified into two categories: type 1 and 2 (\citealt{khachikyan_spectroscopic_1971}, later refined by \citealt{1981ApJ...249..462O}). Type 1 AGN have bluer spectra, strong continuum emission from the accretion disk, and extremely broad emission lines. Type 2 AGN lack continuum emission and broad emission lines, and are instead dominated by narrow-line emission. These types are also called unobscured and obscured, respectively. This refers to the Unified Model of AGN \citep{Antonucci_1993}, which proposed that the difference between type 1 and 2 AGN is obscuration along the observer's line of sight. Emission from the accretion disc and broad line region of the AGN is blocked by an obscuring torus of dust when viewed side-on. However, when viewed face-on the torus does not intervene allowing light from the accretion disc and broad line region (BLR) to reach the observer. This allows for the two distinct populations of AGN to be interpreted as a single phenomenon with features driven by inclination.    

AGN vary across a wide range of timescales. Overall, accretion phases and host galaxy feedback operate over $10^5 - 10^6$ year life cycles \citep{schawinski_active_2015}. However, processes close to the central supermassive black hole can drive variability on human timescales. `Changing Look AGN' (CLAGN) undergo dramatic spectral shifts, losing or gaining broad emission lines and optical continuum emission over mere months to years \citep[e.g.][]{1971ApJ...164L.109K,1976ApJ...210L.117T,1986ApJ...311..135C,McElroy_2016,temple_BASS_2023,Ricci_2023}. Such rapid variations cannot be explained by static orientation-based unification models, making CLAGN crucial laboratories for probing short-term accretion disk instabilities, dynamic obscuration, and complex unknown physics in real-time.

\cite{Ricci_2023} proposed that CLAGN actually contain two distinct populations, Changing-State AGN (CSAGN) and Changing-Obscuration AGN (COAGN). The former transition between observed spectral type due to intrinsic changes in the ionising radiation from the accretion disc \citep{2018MNRAS.480.3898N}. This can cause large changes in the continuum emission of the AGN and cause broad emission lines to appear or disappear -- leading to a change in spectral classification. The latter appear to change spectral type due to varying obscuration along the line of sight between the observer and the core of the AGN. A change in obscuration modulates the amount of continuum emission from the nucleus and broad line emission that reaches the observer leading to a change in the observed spectral type \citep[e.g.][]{Zeltyn_2024}. 

Early discoveries of CLAGN were limited to isolated chance discoveries such as Mrk 1018 and Mrk 590 \citep{1986ApJ...311..135C, McElroy_2016, 2014ApJ...796..134D}. Modern multi-epoch photometric and spectroscopic surveys such as the Sloan Digital Sky Survey \citep[SDSS;][]{2009ApJS..182..543A}, the Dark Energy Spectroscopic Instrument \citep[DESI;][]{2022AJ....164..207D}, and the Zwicky Transient Facility \citep[ZTF;][]{2019PASP..131a8002B} have expanded the cadence of our observations of AGN allowing us to create larger, statistical samples comprising up to hundreds of confirmed CLAGN \citep[e.g][]{2022MNRAS.511...54H,2024MNRAS.535.2322A,2025ApJS..278...28G, chen_DESI_2026}. Within these large statistical samples, a smaller subset have been observed to transition multiple times. These recurring CLAGN (RCLAGN) were also first found in isolated observations \citep[e.g. Mrk 1018;][]{McElroy_2016}, but larger samples of up to $\sim60$ objects are now starting to be found \citep{chen_DESI_2026}. RCLAGN serve as ideal targets for studying the CL phenomena \citep{wang_instability_2024, wang_recurrent_2025}.  Of the currently known RCLAGN, only $\sim10$ are found below $z=0.1$.

\begin{figure}
    \centering
    \includegraphics[width=0.8\linewidth]{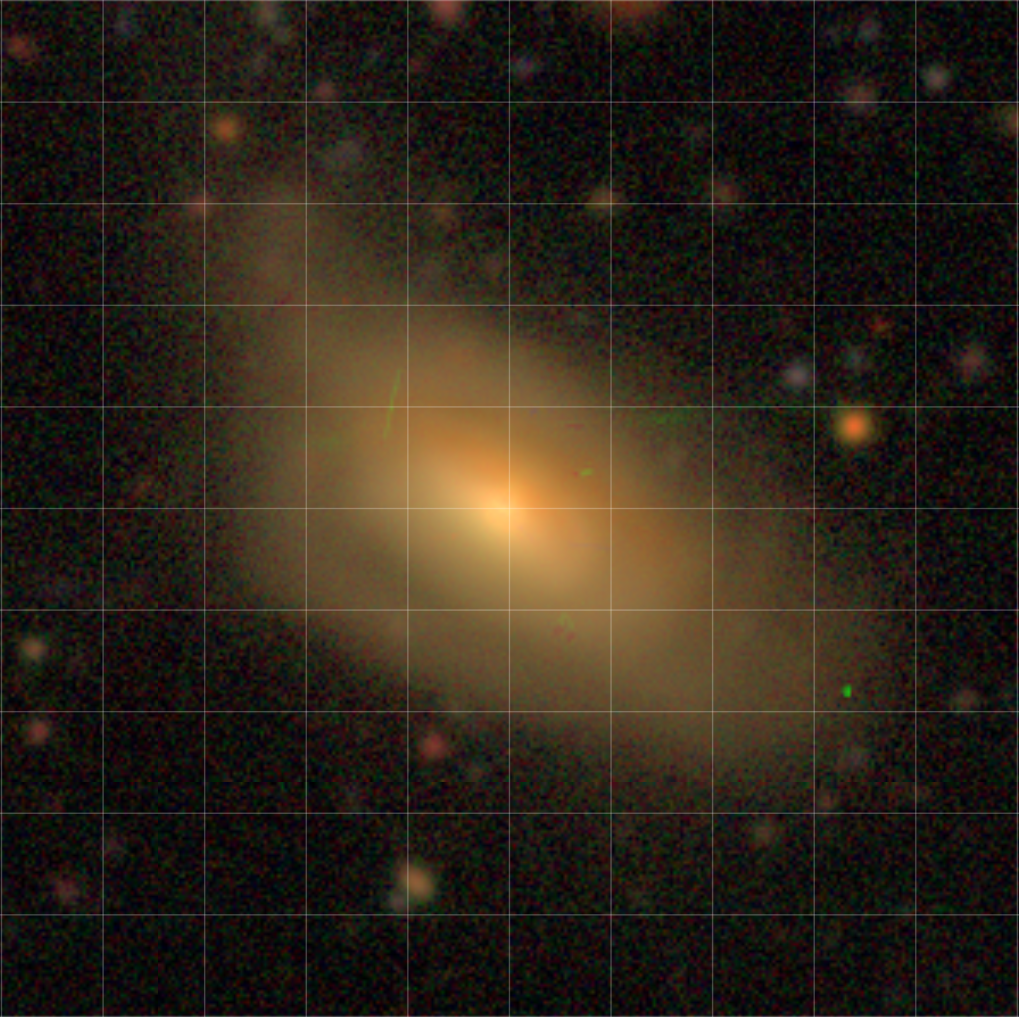}
    \caption{Pseudo-true-colour image of \objname\ based on $gri$ imaging from the Hyper Suprime-Cam Subaru Strategic Program \citep{2022PASJ...74..247A}.  A $50\times50$ arcsec image with grid spacing of 5 arcsec.  There is visible evidence of a central dust-lane and low-surface-brightness features suggestive of the galaxy being post-merger.}
    \label{fig:hsc_image}
\end{figure}

In this paper, we present an analysis of SDSS J085337.26+014303, henceforth \objname, a visually disturbed red disc-like galaxy at $z=0.0582$ (Figure \ref{fig:hsc_image}). It was identified as a changing look AGN based on observations from SDSS and the Sydney AAO Multi-object Integral field spectrograph (SAMI; \citealt{Croom_2021}).  These data suggest recurring changing look events, so that this galaxy has gone through at least one complete cycle.  In this paper, we explore the characteristics of these changes and propose that they are the result of a change of state (i.e.\ varying accretion) in the nucleus of this galaxy. 

In Section \ref{sec:data} we describe the spectroscopic and other data available for \objname.  In Section \ref{sec:spec} we describe our approach and results for spectral analysis.  Our photometric analysis is presented in Section \ref{sec:phot}.  We discuss our results in Section \ref{sec:discussion} and present our conclusions in Section \ref{sec:conc}.  We assume a cosmology with $\Omega_{\rm m}=0.3$, $\Omega_\Lambda=0.7$ and  $H_0=70$\,\kmsmpc.

\section{Data}\label{sec:data}

\begin{figure*}
    \centering
    \includegraphics[width=\linewidth]{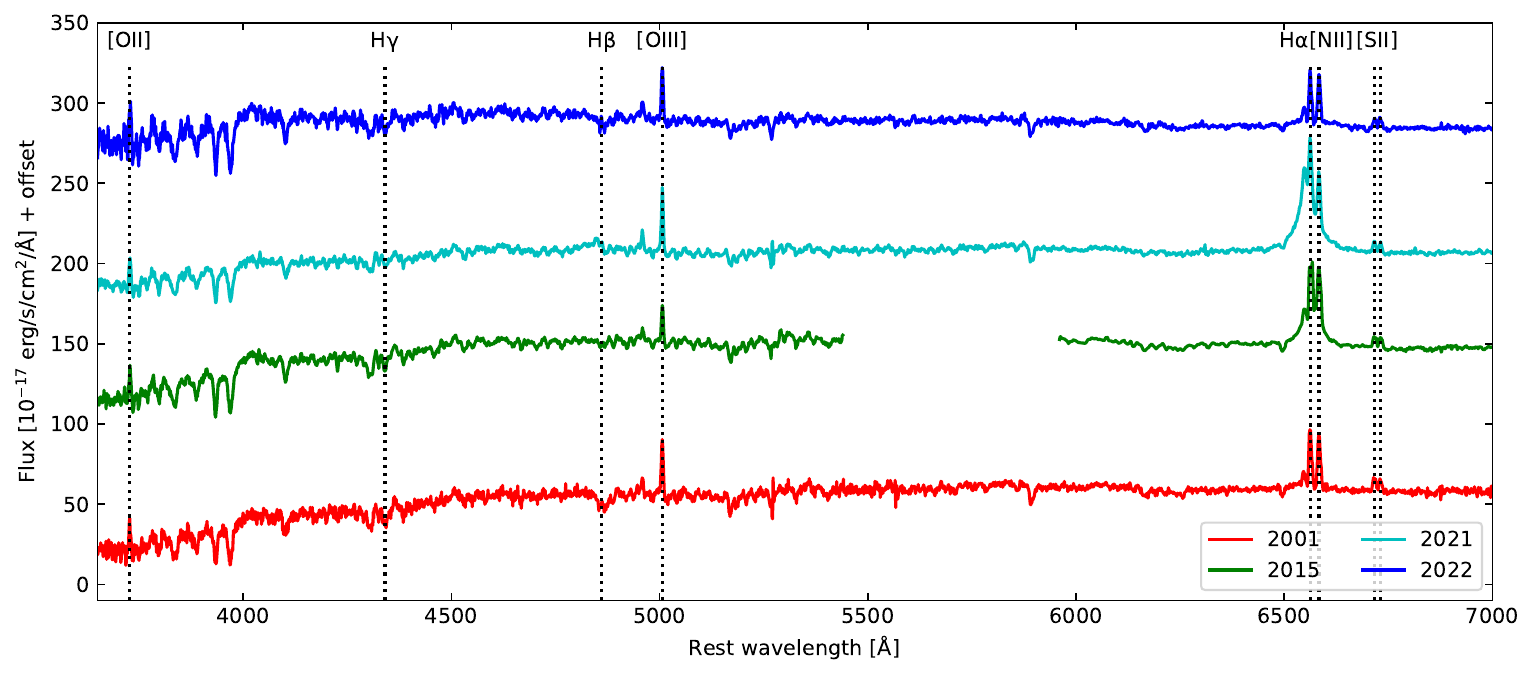}
    \caption{The original spectra (transformed to rest frame) for \objname\ from 2001 (SDSS, red), 2015 (SAMI, green), 2021 (SDSS, cyan) and 2022 (SDSS, blue).The top 3 spectra are offset from their true flux by steps of $80\times10^{-17}$\,\ergscma\ to aid visibility.  Vertical dotted lines show key emission lines.  Broad \ha\ is visible in the 2015 and 2021 spectra.  These epochs also show evidence of infilling of the \hb\ stellar absorption line.  The overall continuum spectral shapes are similar in all cases, although the 2022 spectrum is has slightly bluer continuum, likely caused by flux calibration errors.}
    \label{fig:orig_spec}
\end{figure*}

\objname\ is a galaxy at $z=0.0582$, with a stellar mass of $\log(M_*/M_\odot)=10.7$ \citep{2015MNRAS.447.2857B} and relatively red colours, of $u-r=2.4$ \citep[SDSS model magnitudes;][]{2009ApJS..182..543A} putting it close to the red sequence.  Within the SAMI Galaxy Survey it has an ID number (or CATID) 323854. Deep imaging from the Hyper Suprime-Cam Subaru Strategic Program \citep{2022PASJ...74..247A} shows evidence for low-surface-brightness tidal features, suggestive that the galaxy has previously undergone a merger.  This imaging also suggests the presence of a central dust lane (see Fig.\ \ref{fig:hsc_image}). These features are similar to those seen in the host galaxy of Mrk 1018, as examined extensively in \citep{McElroy_2016}.

\objname\ was first observed spectrally in Jan 2001 (MJD 51913) by the SDSS-I Survey \citep{2009ApJS..182..543A}.  The SDSS-I spectrographs covered a wavelength range of $3800-9200$\,\AA\ at resolution $R\simeq1800-2200$ using 3 arcsec diameter fibres.
The second spectrum was taken in Jan 2015 as part of the SAMI Galaxy Survey, which uses a multiplexed integral field unit (IFU) system on the Anglo-Australian Telescope (AAT) \cite{2012MNRAS.421..872C,Croom_2021}. The AAT's AAOmega spectrograph has two arms, which for SAMI covers $3750-5750$\,\AA\ and $6300-7400$\,\AA\, with resolutions of $R=1808$ and $4304$ respectively. Each SAMI IFU is 15 arcsec in diameter, from which we generated aperture spectra to match SDSS \citep{2011MNRAS.415.2173B,2012MNRAS.421..872C,2015MNRAS.447.2857B}.  As SDSS spectra are calibrated to PSF magnitudes (to correct for flux missing outside of the fibre, assuming a point source), we do the same when generating SAMI aperture spectra, correcting for missing flux, given the seeing during the SAMI observation. We note that this calibration is suitable for the unresolved component in the spectrum (AGN continuum and BLR), but not for the host galaxy.  In our analysis below, we will generally make use of these PSF magnitude calibrated spectra, but we note explicitly where we use other calibrations.

Comparisons between the 2001 SDSS spectrum and the 2015 SAMI spectra identified \objname\ as a potential CLAGN. Further observations as part of SDSS-V \citep{2025arXiv250707093S} in Feb 2021 and Mar 2022 showed continuing variability.  These spectra used the updated SDSS spectrographs \citep{2013AJ....146...32S} with 2 arcsec diameter fibres, wavelength coverage of $3650-9500$\,\AA\ and resolution $R=1560-2650$. \objname\ is also listed as a potential changing look AGN by \citet{2024ApJ...966...85Z} on the basis of the 2001 and 2021 spectra.

We also examine archival photometry of \objname.\ This includes infrared Wide-field Infrared Survey Explorer \cite[WISE]{2010AJ....140.1868W} data taken as part of the Near Earth object WISE project \citep[NEOWISE][]{2011ApJ...731...53M} W1- and W2-band photometry (3.4 and 4.6$\mu$m, using Vega zero-pointed magnitudes). In the optical we make use of various time-domain surveys.  The Catalina Real-Time Transient Survey \citep[CRTS;][]{2009ApJ...696..870D} extends from 2005 to 2016 providing unfiltered photometry transformed to an approximate $V$-band.  
Finally, the Zwicky Transient Facility (ZTF) survey  \citep{2019PASP..131a8002B,masci2023newforcedphotometryservice} has \textit{r}-band photometry from 2018.  We note that across these different imaging surveys, photometry is carried out in various different ways (different apertures, band, PSF etc), so directly calibrating them to a single uniform pass-band is non-trivial, especially for galaxy photometry.  Therefore, we largely choose to consider them separately and look for qualitative consistency across different data sets.

We queried all publicly available survey data from the Australian Square Kilometre Array Pathfinder (ASKAP) to produce a long-term radio light curve spanning October 2019 to April 2026. While the source is within the footprint of the ASKAP Variables And Slow Transients extragalactic survey \citep{2026PASA...43...39D} which has regular observations from June 2023 onwards, it is substantially below the detection threshold. Instead, we are limited to deeper observations obtained as part of the other survey science projects, which generally have minimal repetitions of the field, resulting in an overall sparse light curve. For observations with a catalogued detection, we use the corresponding flux density measurements. There are several observations where the source qualitatively appears in the image but is not present in the catalogue and in these instances we carry out forced photometry\footnote{\url{https://github.com/askap-vast/forced_phot}} at the source location to estimate the flux density.

\section{Spectroscopic  analysis of \objname}\label{sec:spec}

\subsection{Spectral fitting approach}

The four different spectral epochs are displayed in Figure \ref{fig:orig_spec}.  The spectra from 2015 (green) and 2021 (cyan) show evidence of an increased BLR component in the \ha\ line.  These two epochs also show evidence of increased \hb\ emission, with the \hb\ stellar absorption being infilled in 2015 and 2021.  There is no obvious evidence of an AGN-like power-law continuum in any of the spectra, with the continuum appearing to be dominated by host galaxy light in all cases, although we will quantified this in detail below. The most recent spectrum from 2022 (blue) has a continuum shape that is bluer than the other cases, but as we will demonstrate below, this appears to be largely due to errors in relative flux calibration.  There is no significant change in the narrow line region (NLR) emission lines.

To quantitatively analyse the spectra we make use of \ppxf\ \citep{2004PASP..116..138C,2017MNRAS.466..798C,2023MNRAS.526.3273C} to simultaneously fit both the stellar continuum and emission line components.  We use the E-MILES stellar population templates \citep{2016MNRAS.463.3409V} that have a spectral resolution of FWHM = 2.5\AA.  We use a solar-scaled $\alpha$-element abundance, 25 logarithmically spaced ages from 63\,Myr to 15.8\,Gyr and 6 metallicities with $\log(Z/Z_\odot)=-1.71$, $-1.31$, $-0.71$, $-0.4$, $0.0$ and $0.22$.  The SAMI red arm data is convolved to the same resolution as the blue arm (FWHM = 2.66\AA) to enable us to fit red and blue arms simultaneously.  This has no impact on the resulting fitted properties of the BLR gas.

For fitting the gas, we model all significant emission lines visible in the spectrum.  For the SDSS spectra, a single kinematic component is fit for the narrow gas components (i.e. a combination of NLR and general interstellar medium emission).  The \oiii\ $\lambda\lambda$5007,4959 doublet is fit as a separate kinematic component, as this is often broader in AGN. However, in practice it is not found to be significantly broader than the remainder of the narrow lines.  For the SAMI spectrum only we fit a second narrow component, as the higher resolution leads to clear double peaks in the narrow \ha\ line, along with evidence (asymmetric lines) of a second component in \nii\ $\lambda\lambda$6548, 6583 and \sii\ $\lambda\lambda$6716, 6731.  For the broad lines, we fit between 1 and 3 Gaussian components, depending on the strength of the broad lines, only increasing the number of components if it made a clear improvement in quality of the fit (examining the fit visually and comparing reduced $\chi^2$).  The Balmer line components are tied together kinematically, as this allows us to place reasonable constraints on the Balmer decrement, even in the case of weak or undetected \hb\ emission.  For the 2001 and 2022 spectra, only 1 broad component was fit, while for the 2015 and 2021 spectra we fit 3 broad components.

One of the challenges of fitting single-fibre data, such as that from SDSS, is that while overall flux calibration is good, individual spectra can have differences in spectral shape due to flux calibration. These differences can be large enough to influence the overall fit, particularly for later spectra using the smaller 2 arcsec fibres \citep[e.g.][]{2016ApJ...831..157M}.  While the flux calibration is typically better for IFS data given the reduced aperture losses, there is a small average difference in the spectra shapes of SDSS and SAMI data which is up to $\sim10$ percent at 4000\,\AA\ \citep{Croom_2021}.  A further concern is that SDSS spectra are normalised to PSF magnitudes of point sources in the field.  This normalisation should correct the unresolved BLR flux to a total value, but will underestimate the total host galaxy flux.  We therefore use a multiplicative polynomial for our fitting in \ppxf, with {\tt mdegree} of 10.  Across the wavelength range of interest ($\sim4500-7000$\,\AA\ in the rest frame), the multiplicative polynomial varies between $\sim0.9$ and 1.1 for all spectra, so the corrections that this applies are modest.  We also re-run the fits with an additional additive polynomial with {\tt degree} of 6.  This allows us to explore whether there is any evidence of an AGN continuum from the accretion disk in our spectra.

One of our parameters of interest is the dust extinction of the BLR.  To account for this, we allow for two separate gas dust models, one for the broad lines and a second for the narrow lines.  We do not apply a dust model to the stellar population templates, as we already use a global multiplicative polynomial.  The dust model for the gas follows a \citet{2000ApJ...533..682C} extinction law.  As \ppxf\ multiplies templates by both the multiplicative polynomial and the attenuation curve, we estimate the dust obscuration from the Balmer decrement in the full broad line model (with both terms applied).  This is equivalent to assuming there is no flux calibration error in the relative spectral shapes.   To explore the impact of systematic flux calibration errors, we estimate the Balmer decrement in two other ways.  First, we assume that the multiplicative polynomial is purely fixing flux calibration errors, and so take the results of the \ppxf\ gas attenuation directly.  Second, we determine a multiplicative polynomial correction that matches the shape of the SDSS stellar continuum to the SAMI spectrum.  We then recalculate the Balmer decrement from fits to these re-calibrated spectra.  Our conclusions are not qualitatively changed when we calculate the extinction in these different ways.

Uncertainties on the \ppxf\ fits are estimated using a bootstrapping approach.  We add the residuals from the best fit onto the best fit model and bootstrap this 1000 times, using a wild bootstrap approach \citep[e.g.][]{DAVIDSON2008162}.  All the uncertainties related to \ppxf\ fitting are quoted as the $1\sigma$ (16th-84th percentile) uncertainties from our bootstrapping.

\subsection{Evidence of AGN continuum}

We first test whether there is evidence of AGN continuum in any of the spectra.  To do this, we measure the \dnfour\ index \citep{1999ApJ...527...54B}, and use the seeing corrected 3 arcsec diameter SAMI aperture spectrum to consistently compare to SDSS.  The addition of a power-law AGN continuum should lead to a change in the index measurements between epochs.  We make the index measurements on the spectra after subtracting the best fit emission lines (broad and narrow) from the original spectrum and adjusting the stellar continuum shape to match the SAMI 2015 spectrum.  The continuum correction is done to correct for potential flux calibration differences (particularly in the 2022 spectrum).  Such a correction might change the AGN continuum estimate, but given that the AGN continuum contribution is weak (see below), correcting the flux calibration differences gives us a more robust estimate of \dnfour.  The \dnfour\ index values are  $1.569\pm0.012$, $1.502\pm0.004$, $1.487\pm0.006$ and $1.546\pm0.010$ for 2001, 2015, 2021 and 2022 respectively.  The uncertainties quoted are purely statistical, propagated from the variances in the individual spectral pixels.  We note that the 2015 and 2021 spectra, where the broad lines are clearly visible, have the lowest \dnfour, suggestive of a non-stellar continuum in these spectra.  The difference is consistent with adding an AGN continuum of $\sim10-15$ percent of the stellar flux at 4000\AA. 

As a second test for continuum emission we use the \ppxf\ fits including an additive polynomial term ({\tt degree}=6).    We calculate the amplitude of the best fit additive continuum at 4000\,\AA\ (averaged within a 100\,\AA\ window).  Uncertainties are estimated from the bootstrap resampled \ppxf\ fits.  The fractional contribution of the additive polynomial to the total continuum, $f_c$, is given in Table \ref{tab:spec_results} for 4000\AA.  For 2001 and 2022, the extra continuum is consistent with zero, while for 2015 and 2021 it contributes $0.137\pm0.010$ and $0.102\pm0.012$ of the flux.  These values are consistent with the \dnfour\ analysis above and confirms the presence of weak, but detectable, AGN continuum when we see obvious broad lines.   Given degeneracies with stellar population residuals, flux calibration uncertainty and dust, as well as the low level of AGN continuum, it is hard to accurately interpret the shape of this continuum component.  However, we estimate the best power-law, $f_\lambda\propto\lambda^\beta$, across the wavelength range $4000-7000$\,\AA\ in the rest frame by fitting to the additive polynomial.  We find $\beta=-0.4\pm0.8$ and $+2.2\pm0.6$ for the 2015 and 2021 spectra respectively (uncertainties from bootstrapping).  While these are different from each other, they are both considerably redder than typical AGN continua that have $f_\lambda\sim\lambda^{-1.5}$ or equivalently, $f_\nu\sim\nu^{-0.5}$ \citep[e.g.][]{2001AJ....122..549V}. This suggests that the continuum may be partially dust obscured.  However, we do not formally estimate obscuration for this continuum given the difficulty of robustly fitting this.

Finally, we note that although the low-state SDSS spectrum from 2022 suggests a bluer spectral shape (Figure \ref{fig:orig_spec}), the lack of evidence for any AGN continuum in this spectrum confirms our suspicion that the shape is largely related to uncertainty in flux calibration.  Once a correction is made to the flux calibration, stellar absorption features are consistent between the 2001 and 2022 spectra.

\begin{table*}[ht]
    \centering
    \setlength{\tabcolsep}{4pt} 
    \begin{tabular}{lllccccccccc}
    \toprule

MJD & Date & Source & $f_c$(4000\AA) & $E(B-V)$ & $\Delta v_{BLR}$ & FWHM$_{BLR}$ & $\sigma_{BLR}$ & $F_{H\alpha,BLR}$ & $L_{H\alpha,BLR}$ & $L_{H\alpha,BLR,int}$ & $M_{BH}$\\
    &       &    & & mag. & \kms\ & \kms\ & \kms\ & $10^{-15}$\ergscm\ & $10^{40}$\ergs\ & $\log$(\ergs) & $10^6$\msun\ \\
\hline
51913 & 04/01/2001 & SDSS & $-0.019\pm0.023$ &  $7.52^{+2.21}_{-2.63}$ & $78\pm169$ & $5847\pm514$ & $2482\pm215$ & $4.4\pm0.4$ & $3.6\pm0.3$ & $42.6-47.5$ & $12.2\pm3.3$ \\
57047 & 25/01/2015 & SAMI & $0.137\pm0.010$ &  $0.55^{+0.05}_{-0.04}$ & $166\pm15$ & $2232\pm84$ & $1040\pm54$ & $12.9\pm0.3$ & $10.4\pm0.3$ & $41.58\pm0.05$ & $3.0\pm0.6$ \\
59252 & 07/02/2021 & SDSS & $0.102\pm0.012$ &  $0.70^{+0.04}_{-0.04}$ & $-186\pm22$ & $2332\pm66$ & $1575\pm55$ & $26.8\pm0.6$ & $21.7\pm0.5$ & $42.04\pm0.04$ & $4.9\pm0.9$ \\
59664 & 26/03/2022 & SDSS & $-0.003\pm0.018$ &  $5.92^{+1.89}_{-2.08}$ & $477\pm68$ & $4793\pm210$ & $2035\pm89$ & $5.0\pm0.2$ & $4.0\pm0.2$ & $42.7-46.8$ & $8.6\pm1.7$ \\
    \toprule \\
    \end{tabular}
    \caption{Spectroscopic results for \objname.  We list the key results from spectral fitting with uncertainties derived from bootstrap estimates.  Results for SAMI are based on 3 arcsec aperture spectra, corrected for seeing loss (to be consistent with SDSS).  We provide results for: $f_c(4000\AA)$, the fraction of non-stellar continuum at 4000\AA; $E(B-V)$; $\Delta v_{BLR}$, the offset from the stellar continuum rest frame of the \ha\ BLR emission; FWHM$_{BLR}$ and $\sigma_{BLR}$, the FWHM and standard deviation of the BLR (measured from the BLR \ha); $F_{H\alpha,BLR}$, the BLR \ha\ flux without extinction correction; $L_{H\alpha,BLR}$, the BLR \ha\ luminosity without extinction correction; $L_{H\alpha,BLR,int}$ the BLR \ha\ luminosity after extinction correction; and $M_{BH}$, the black hole mass estimated from the \citet{Greene_2005} virial relation (using the measured luminosity without extinction correction).  For the low state observations we only provide a $1\sigma$ range for $L_{H\alpha,BLR,int}$.}
    \label{tab:spec_results}
\end{table*}

\begin{figure*}
    \centering
    \includegraphics[width=\linewidth]{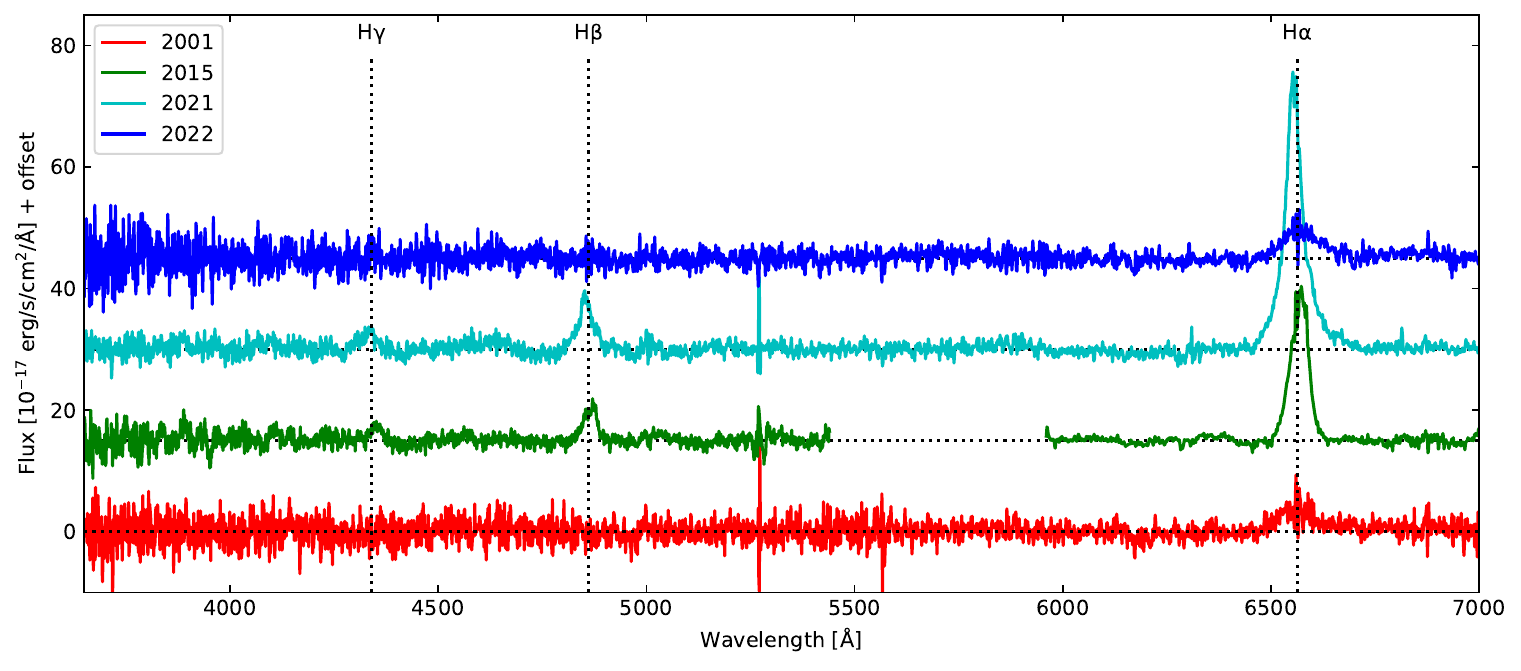}
    \caption{The host (stellar continuum and narrow emission lines) subtracted spectra for \objname\ from 2001 (SDSS, red), 2015 (SAMI, green), 2021 (SDSS, cyan) and 2022 (SDSS, blue).The top 3 spectra are offset from their true flux by steps of $15\times10^{-17}$\,\ergscma\ to aid visibility.  Vertical dotted lines show key emission lines in the rest frame.  Broad \ha\ is visible in all spectra, but is weak in 2001 and 2022.  Higher order Balmer lines are only visible in 2015 and 2022.}
    \label{fig:hostsub_spec}
\end{figure*}

\begin{figure}
    \centering
    \includegraphics[width=\linewidth]{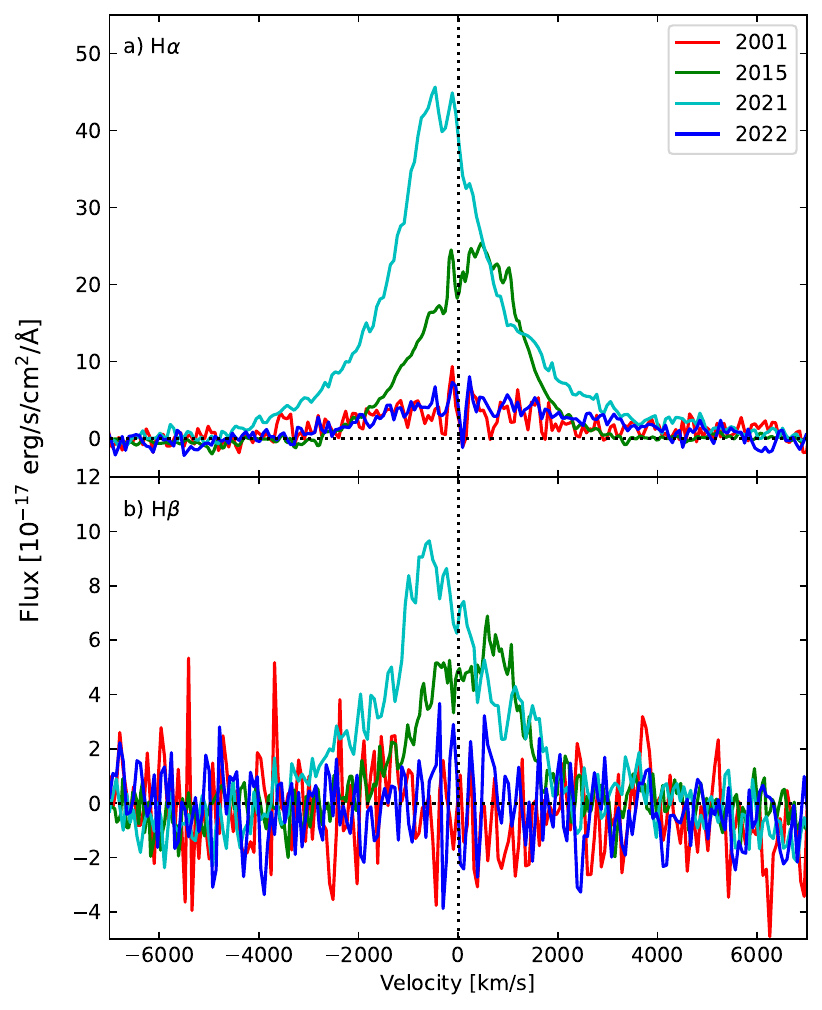}
    \caption{A zoom-in view of the a) \ha\ and b) \hb\ broad lines for the host subtracted spectra for \objname\ from 2001 (SDSS, red), 2015 (SAMI, green), 2021 (SDSS, cyan) and 2022 (SDSS, blue).  Velocities are with respect to the rest frame defined by the stellar continuum.}
    \label{fig:blr_zoom}
\end{figure}

\subsection{Properties of the broad-line emission}

The most obvious difference between different epochs is the presence of strong broad \ha\ emission in 2015 and 2021.  To more clearly visualise this, we subtract the continuum and narrow line gas from the \ppxf\ fits, to show a broad-line only spectrum in Figure \ref{fig:hostsub_spec}.  The spectra in 2001 and 2022 show weak but significant broad \ha, but there is no evidence of higher order Balmer emission.  In contrast, 2015 and 2021 show much stronger \ha\ emission and also significant \hb\ and H$\gamma$.  The 2021 spectrum in Figure \ref{fig:hostsub_spec} also shows a possible broad hump between the \hb\ and H$\gamma$ lines that would be consistent with broad Fe{\small II} emission \citep[e.g.][]{2001ApJS..134....1V}, although given the weakness of this feature we do not attempt to fit it directly.  In the high state, the flux ratio of broad \hb\ to narrow \oiii\ is \hb/\oiii$\simeq1.6-1.8$, which would put this object in the Seyfert 1.2 category \citep[e.g.][]{2016MNRAS.462.3570S}.

To confirm that \objname\ would have been detected as a CLAGN in other surveys we calculate the broad line flux ratio, 
\begin{equation}
R=(S_b-S_d)/S_b,
\label{eq:fluxratio}
\end{equation}
where $S_b$ and $S_d$ are the fluxes of the broad line in the bright and dim state respectively \citep{2025ApJ...986..160D}.  We find $R_{H\alpha}=0.81\pm0.01$ and $R_{H\beta}=1.00\pm0.03$ between 2021 and 2022 (and similar values between other epochs), substantially higher than the $R>0.3$ or $0.4$ threshold often used for CLAGN selection.

To quantify potential obscuration we measure the Balmer decrement in the broad-line residual spectra (Figure \ref{fig:hostsub_spec}).  We assume a \citet{2000ApJ...533..682C} extinction law, but assume that the intrinsic Balmer decrement ratio is $R=3.1$, rather than the standard case-B recombination value of 2.86, as this seems to more closely approximate the true ratio in broad-line regions \citep[e.g.][]{1984PASP...96..393G,2008MNRAS.383..581D}.  While the low scatter in Balmer decrement found in blue quasars by \citet{2008MNRAS.383..581D} suggests that they can be reliably used for extinction estimates, we note that BLR gas that is optically thick to Balmer emission \citep[e.g][]{RevModPhys.51.715,2016MNRAS.462.3570S} can lead to intrinsic variations in Balmer decrement.  For example, the model grid of \citet{2016MNRAS.462.3570S} spans a range of intrinsic \ha/\hb\,$\sim2.85-6.5$.  However, the range of intrinsic Balmer decrements may be narrowed through a locally optimally-emitting cloud model \citep{1995ApJ...455L.119B}, with \citet{2008MNRAS.383..581D} finding a scatter of only 0.03 dex.  Below we give statistical errors on our $E(B-V)$ estimates, but note that larger systematic uncertainties are possible.

Both the 2015 and 2021 spectra show moderate extinction with $E(B-V)=0.55^{+0.05}_{-0.04}$ and $0.70\pm0.04$ respectively (see Table \ref{tab:spec_results}).  Given that the \ha\ line flux is greater in 2021 compared to 2015 (by a factor of $\sim2$, see table \ref{tab:spec_results}), this means that the intrinsic (extinction corrected) \ha\ broad-line luminosities are not consistent.  The measured Balmer decrements are $5.37\pm0.24$ (2015) and $6.18\pm0.23$ (2021), which are at the high end of plausible intrinsic optically thick models, so moderate dust obscuration does seem more likely, particularly given the visible dust lane in Figure \ref{fig:hsc_image}.

The measured dust attenuation from the 2001 and 2022 spectra is considerably higher than 2015 and 2021.  There are larger uncertainties, due to the combination of weaker \ha\ and non-detection of \hb, but these measurements still allow a reasonable estimate of $E(B-V)$, with values of  $7.52^{+2.21}_{-2.63}$ and $5.92^{+1.89}_{-2.08}$ for 2001 and 2022 respectively.  The $2\sigma$ lower limits on the Balmer decrement for these two epochs are 12.5 (2001) and 15.9 (2022).  Both are substantially higher than reasonable intrinsic optically thick ratios.
This demonstrates that BLR in the low-state spectra has significantly higher dust extinction than the high-state spectra.  Given the high extinction, with large uncertainty, the intrinsic luminosity of \ha\ in the low-state BLR ($L_{H\alpha,BLR,int}$) is highly uncertain (by orders of magnitude, see Table \ref{tab:spec_results}). 

We now consider the BLR gas kinematics in greater detail.  First, we note that there are clear kinematic differences between epochs.  Figure \ref{fig:blr_zoom} shows the \ha\ and \hb\ line profiles on a velocity scale.  The two high-state spectra show a significant offset in their mean velocity.  We calculate the velocity centroid of the BLR with respect to the rest frame (defined by the stellar population) and find that this $\Delta v_{BLR}=166\pm15$ and $-186\pm22$\,\kms\ for the 2015 and 2021 spectra respectively.  This cannot simply be a case of the entire BLR increasing and decreasing in brightness.  The velocity shift for the weak 2022 \ha\ BLR is also significantly different from other epochs with $\Delta v_{BLR}=447\pm68$\,\kms, while the uncertainties on $\Delta v_{BLR}$ are larger for 2001.

We measure the FWHM and 2nd moment (i.e. $\sigma_{BLR}$) of the BLR based on the full BLR model fit using \ppxf\ (with bootstrap based uncertainties, see table \ref{tab:spec_results}).  The FWHM measurements of the two high-state spectra are consistent with each other, while the $\sigma_{BLR}$ values are significantly different (the 2021 spectrum has a larger $\sigma_{BLR}$), suggestive of broader wings in the 2021 line profile, that can be seen in Figure \ref{fig:blr_zoom}a.  The two low-state spectra both have broader profiles (as measured by FWHM and $\sigma_{BLR}$) than the high state spectra.  This difference is significant, despite the relatively low S/N of the broad component in the low state spectra.

\citep{Greene_2005} provide a scaling relation to estimate $\mbh$ from just \ha\ line width and luminosity.   This relationship is  
\begin{multline}
    \mbh = (2.0^{+0.4}_{-0.3})\times10^6\left(\frac{L_{H\alpha}}{10^{42}\text{erg}\,\text{s}^{-1}}\right)^{0.55\pm0.02} \\
    \times \left(\frac{\text{FWHM}_{H\alpha}}{10^3\,\text{km}\,\text{s}^{-1}}\right)^{2.06\pm0.06} M_{\odot}.
    \label{eq:green_bhmass}
\end{multline}
We can use this relation to estimate $\mbh$ for \objname\ and the values are given in the last column of Table \ref{tab:spec_results}.  Given the large variation in Balmer decrement (and also the large uncertainties in the obscuration for the low-state spectra), we choose to estimate $\mbh$ using an \ha\ luminosity that is not corrected for obscuration, noting that this may underestimate the true value.  The high-state spectra give $\mbh$ values that are similar, $\mbh\sim3-5\times10^6$\,\msun, although formally inconsistent with each other at the $\sim1.7\sigma$ level.  Applying extinction correction to $L_{H\alpha,BLR}$ would only make this discrepancy worse.  Despite being lower in luminosity, the low-state spectra lead to a higher $\mbh$ estimate, due to their larger FWHM.  Correcting for obscuration would lead to these values being even larger.  It is clear from the $\mbh$ calculation that different epochs do not provide consistent values.  There are several explanations for this, for example, if we are viewing different parts of the BLR, the geometric factor in the Keplerian relation may be different between epochs.  This differing view may be due to intrinsic variation, or changing obscuration, with the second of these being the simplest explanation given that we see changes in $E(B-V)$.  Alternatively, the system is simply not in dynamical equilibrium and so scaling relations under this assumption provide a poor estimate of $\mbh$.

Given the stellar mass of \objname\ of $\log(M_*/M_\odot)=10.7$, we can examine whether the BH mass is consistent with known scaling relations (assuming our high state $M_{BH}$ values).  The BH mass we calculate is somewhat below the mean scaling relations found by \citet{2023MNRAS.518.2177G}, but within the scatter of the population (see their figure 5).  In particular, the BH mass is relatively typical for spirals in this stellar mass range. If we calculated the BH mass after extinction correction, the mass would be closer to the mean $M_{BH}-M_*$ relation.  

As an independent analysis of the BLR of \objname\ we take difference spectra, subtracting the 2001 spectrum from the others.  While this will subtract the low-level broad component seen in our low-state spectra, it is useful as it provides an analysis independent of our \ppxf\ fitting.  Before doing this, we adjust the flux calibration of the spectra so that the stellar continuum shape matches (normalising all the spectra to the SAMI 2015 spectrum).  This analysis is only valid for the high-state spectra.    Our difference spectra result in BLRs that have velocity offsets, Balmer decrements and FWHMs that are consistent with our \ppxf\ analysis above.  This difference spectrum test also demonstrates that the NL components do not change significantly between epochs, as expected.

\begin{figure*}
    \centering
    \includegraphics[width=0.85\linewidth]{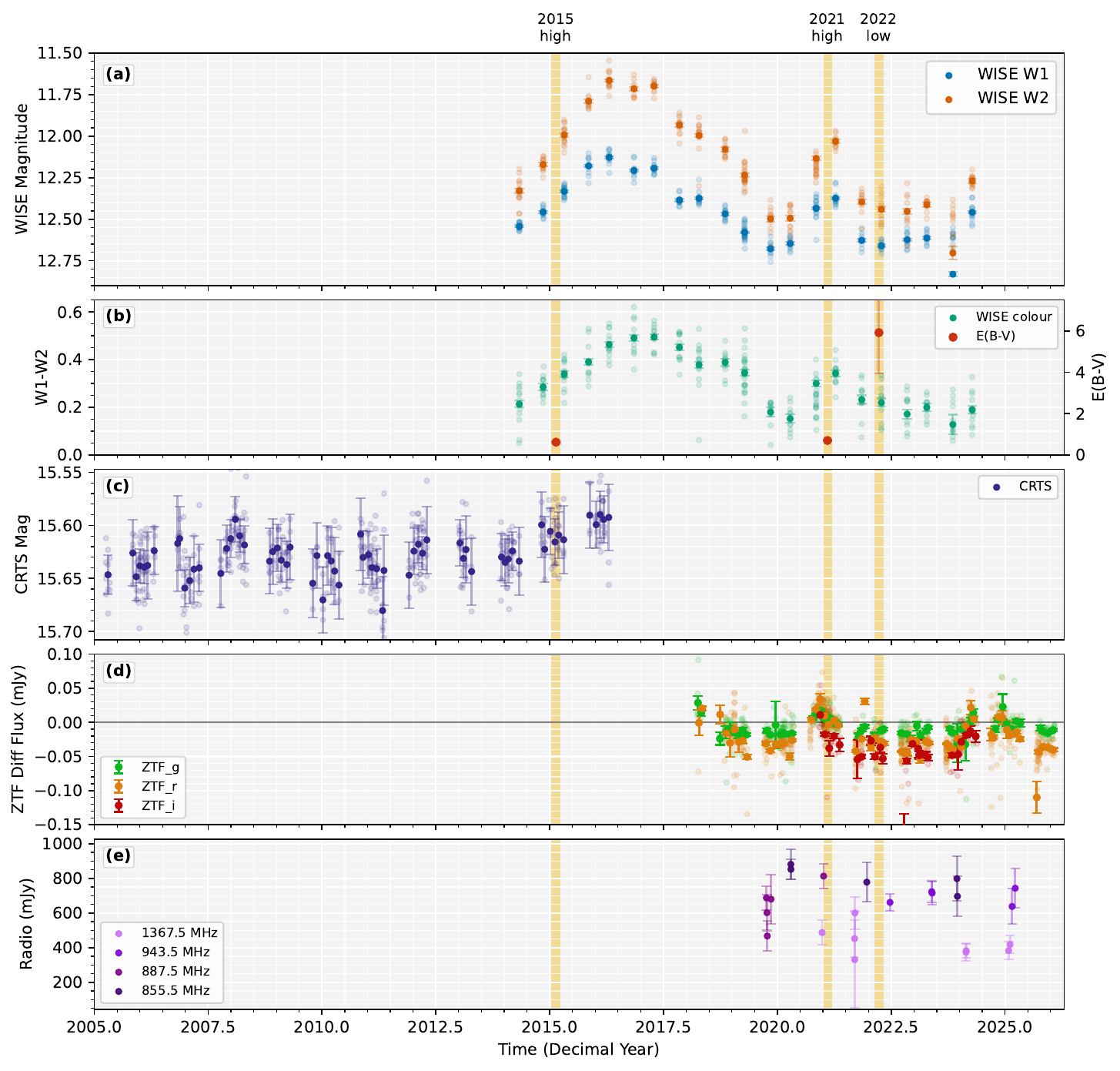}
    \caption{Time-series photometry for \objname.  From top to bottom we show: WISE W1 and W2  magnitudes, WISE colours \& $E(B-V)$ derived from the Balmer decrement of the spectra, CRTS photometry, ZTF difference photometry, and radio flux from ASKAP. Vertical yellow lines show when spectra were taken, and are labelled as 'high' or 'low' state. In all panels individual photometric measurements are shown by transparent symbols, while the opaque symbols with error bars show the mean in 0.1 year intervals.}
    \label{fig:phot}
\end{figure*}

\section{Photometric light curves}\label{sec:phot}

Photometric measurements of \objname\ are shown in Fig.\ \ref{fig:phot}.  First, we examine time-series measurements from NEOWISE \citep{2011ApJ...731...53M}.  Note that all WISE photometric magnitudes are quoted on the Vega scale.  These data start in 2014, so do not cover the low state in 2001.  However, the NEOWISE data show a clear symmetric rise and fall, peaking in brightness in 2016.  There is also a second lower amplitude peak in 2021.  The first high-state spectrum is observed during the first WISE rise in brightness, while the second high state spectrum is observed during the second, briefer rise in 2021.  By the time the low-state spectrum is observed in 2022, the NEOWISE photometry has declined again.  We can also calculate the time evolution of the WISE W1-W2 colour (see Fig.\ \ref{fig:phot}).  In the low state, $W1-W2\simeq0.2$\,mag (for example, the WISE observation closest in time to the 2022 low-state spectrum has $W1-W2=0.219\pm0.018$, which is relatively typical for non-active galaxies.  However, in the bright high state, the colour is redder, with $W1-W2\simeq0.5$ (the reddest colour being the first observation in 2017, with $W1-W2=0.494\pm0.011$).

The original WISE imaging survey was carried out in 2011.  Based on those data, \citet{2014ApJ...782...90C} derived WISE photometry for all sources in the Galaxy And Mass Assembly \citep[GAMA][]{2011MNRAS.413..971D} Survey that helps form the input catalogue for SAMI.  The 2011 WISE photometry is consistent with \objname\ being in a low state with $W1=12.44\pm0.02$, $W2=12.28\pm0.02$, $W3=9.25\pm0.04$, $W4=7.72\pm0.24$\,mag, while colours are $W1-W2=0.16$, $W2-W3=3.03$ and $W3-W4=1.53$.  These colours, place \objname\ approximately in the star-forming galaxy region of the WISE colour-colour space \citep[e.g.][]{2020ApJ...903...91Y}.  However, the SAMI IFS data does not show any direct evidence of star formation.  All spaxels with significant emission lines sit within the composite or AGN part of standard ionisation diagnostic diagrams \cite[e.g.][]{1981PASP...93....5B}.  It may be that there is residual star formation in the nucleus that is masked by the AGN emission, or that the visible dust lane (Figure \ref{fig:hsc_image}) is contributing to the WISE colours.  In the high state with a redder colour, the boundaries defined by \citet{2020ApJ...903...91Y} would likely classify this object as an AGN.  

We will choose the first WISE observation of 2022, which is closest to the 2022 low-state spectrum, to be representative of the photometry purely from the host galaxy ($W2_{\rm host}=12.440\pm0.015$\,mag).  The brightest W2 measurement gives $W2_{\rm HOST+AGN}=11.664\pm0.008$\,mag (first WISE observation of 2016).  From these we can separately calculate the brightness of the AGN only component, finding $W2_{\rm AGN}=12.39\pm0.02$\,mag.  Thus, in the W2 band, the AGN peaks at a luminosity that is similar to the entire host galaxy.

\objname\ also has optical time-domain photometry from a number of different data sets.  No single data set covers the entire time-interval in question, but a combination of data samples from 2005 to the present day.  It is worth noting that these imaging surveys have different depths, filter combinations and processing (e.g.\ different photometric apertures).  CRTS shows an upwards trend broadly consistent with the main rise in WISE.  We use ZTF difference photometry (Figure \ref{fig:phot}d) in $g-$, $r-$ and $i-$bands.  As the reference images are constructed from stacks across 2018 (while the object was coming down from a high state), the difference photometry is largely negative.  However, in ZTF we do see an increase in brightness in all three bands around the second weaker peak in 2021. Although not shown, we do investigate the Asteroid Terrestrial-impact Last Alert System photometry \citep[ATLAS;][]{Tonry_2018}.  However, the changes in PSF through this survey dominate any true changes in galaxy brightness, while the public ATLAS difference imaging has different reference images as a function of epoch, making the light curve hard to interpret.

\objname\ is detected at $\sim0.75$\,mJy  at radio frequencies around $\sim1$\,GHz.  For a given frequency, there is no evidence of variation across the time period sampled (Figure \ref{fig:phot}), which includes the period from 2021 to 2022 when the optical spectrum drops from a high state to low state.   Although the frequency range is not large ($0.85 - 1.4$\,GHz), we are able to fit a power law of the form $f_\nu\propto\nu^\alpha$, finding $\alpha=-1.36\pm0.14$.  This suggests a peak at lower frequency and ageing of the synchrotron source electrons \cite[e.g.][]{1962SvA.....6..317K,2000AJ....119.1111B,2017MNRAS.466.2888H}.  Given this spectral index, it is unsurprising that we see no obvious radio variability, as the radio emission is likely to come from an older more extended region (although still unresolved at the resolution of ASKAP).  Much greater radio variability is found in compact flat spectrum sources \citep[e.g.][]{2011ApJS..194...29R}.  The spectral index is also steeper than typically due to star formation \citep{1992ARA&A..30..575C}, suggesting that \objname\ has relatively long term AGN activity.

\section{Discussion}\label{sec:discussion}

The optical spectral evolution of \objname\ demonstrates a change from Type 1.9 (2001) to Type 1 (2015, 2021) and then back to Type 1.9 (2022).  We originally identified \objname\ as a changing-look AGN based on SAMI spectra taken in 2015, where it is one of only 11 objects with visible broad lines in the SAMI Galaxy Survey \citep{Croom_2021}.  \objname\ was first published as a potential changing-look AGN, using the SDSS spectra from 2001 and 2021 \citep{2024ApJ...966...85Z}.  \objname\ therefore becomes one of a growing number of CLAGN that have gone through a complete cycle.  Canonical objects such as Mrk 1018 \citep{1986ApJ...311..135C,McElroy_2016} and Mrk 590 \citep{2014ApJ...796..134D,2018ApJ...866..123M},were among the first to be found, but focused surveys are now finding several tens of objects in this category \citep[e.g.][]{2025ApJ...981..129W,2025ApJ...986..160D,2025arXiv251018445D}.  Only $\sim10$ of these are found below $z=0.1$, and it is worth noting that the flux changes we find are much greater than the nominal $R\sim0.3-0.4$ used to select samples of CLAGN.  \objname\ shows large changes in both \ha\ and \hb, while recent work \citep{2026arXiv260715564Z} seems to suggest this change in both lines is relatively rare (although the reason for this is still uncertain and may be S/N related).

Given the WISE photometry that shows two peaks (a first broad peak, of width $\sim5-6$ years, and a second narrower peak, of width $\sim1$ year) with a minimum around 2020, it seems plausible that \objname\ has actually undergone two complete cycles, but we do not have the spectra in 2020 to formally confirm this.  Optical time-series photometry, although noisy, also broadly agrees with the WISE changes, suggesting the variation extends into the optical continuum.  This is confirmed by weak, but significant, detection of AGN continuum in the high-state spectra.



Considering whether \objname\ is a CSAGN or a COAGN \citep[e.g.][]{Ricci_2023}, we can examine both the spectroscopy and photometry.  The most compelling evidence for a CSAGN comes from the strong variation in WISE photometry.  Emission in the WISE bands should come largely from hot dust in the torus surrounding the central black hole and accretion disk.  A change purely due to obscuration would lead to a larger change at bluer wavelengths, e.g.\ optical continuum.  However, the largest continuum change is seen in WISE.  WISE colours also become redder (from $W1-W2=0.219\pm0.018$ in the low state, to $W1-W2=0.494\pm0.011$ in the high state), moving from a colour typical of regular galaxies, to one that is more common for AGN (particularly noting that the WISE photometry still includes a host galaxy contribution).  By comparison, the change in the optical continuum is significantly less.  A key point here is that the majority of AGN emission in the WISE bands should be coming from bluer light that has been reprocessed through dust absorption and re-emission in the torus.    If the flux in the WISE bands was also absorbed in the low state, this would be re-emitted at longer wavelengths, leading to redder infrared colours.  In contrast, the low state $W3-W4=1.53$ is more similar to regular galaxies \citep[e.g.][]{2017ApJ...850...68C}, rather than heavily obscured galaxies (e.g.\ ultra luminous infrared galaxies or Compton thick AGN).  Further, \objname\ is within the sample of GAMA galaxies fit with host+AGN SEDs by \citet{2022MNRAS.509.4940T}.  They find an AGN fraction of only 0.0003 between 5 and 20$\mu$m, using original WISE photometry from 2011 (likely in the low state).


The two high-state spectra also have \ha\ flux that is inconsistent with a simple COAGN interpretation.  The 2015 SAMI high-state spectrum has observed flux a factor of $\simeq2$ lower than the 2021 SDSS high-state spectrum. However, the 2021 spectrum shows a slightly higher $E(B-V)$ from the Balmer decrement. The BLR shapes also change between the high and low states, and between the two high states.  For example, there is a velocity offset of $\simeq350$\,\kms\ between the line centres for the two high state spectra.  These differences in line profile rule out a simple screen-like obscuration.  The source of the extinction is unclear.  This could be related to dust around the central source (e.g.\ a clumpy torus), but may also be from galaxy scale dust.  The optical image of \objname\ (Figure \ref{fig:hsc_image}) shows evidence of a dust lane that could provide such obscuration.

Given that \objname\ appears to be a CSAGN, we can use the changes in the BLR to try to better understand the nature of the state change.  We can first consider whether the changes we see are consistent with a simple BLR following a virial relation, 
\begin{equation}
    M_{BH} \simeq f\frac{R_{BLR} V^2}{G}, 
\end{equation}
were $R_{BLR}$ is the radius of the BLR, $V$ is the velocity (e.g.\ parameterised by the BLR FWHM), $G$ is the gravitational constant and $f$ is a virial factor that accounts for unknown geometry.  A BLR radius-luminosity relation of the form $R_{BLR}\propto L^{0.5}$, then implies that $V\propto L^{-0.25}$ under the assumption that $f$ and $M_{BH}$ are fixed.  The largest change in FWHM (between 2001 and 2015) is a factor of $2.59\pm0.24$.  This would infer a decline in luminosity by a factor of $0.02\pm0.008$.  This is a much greater change than the observed difference in flux in the \ha\ line (see Table \ref{tab:spec_results}) which is only $0.27\pm0.02$.  If we correct for the larger extinction in the low state, the discrepancy becomes even larger.  Thus, the data are not consistent with this simple model and a more complex physical picture is needed.

A more natural explanation is that the low-luminosity BLR component seen in the low state is coming from a distinct physical source compared to the narrower BLR in the high state.  It is worth noting that the spectra are consistent with the broad low-state BLR component still being present in the high-state spectra (see Figure \ref{fig:blr_zoom}).  It now seems likely that within the general AGN population the BLR is not a single uniform structure, but made up of several components.  Line profiles are often found to be complex, and unable to be fit by simple analytic models.   A number of studies identify reverberation mapping lags that vary as a function of velocity \citep[e.g.][]{2003A&A...407..461K,2016ApJ...820...27D,2020ApJ...905...75H,2022ApJ...925...52U}.  There is no simple picture from such lags.  Some objects show a lower lag for the broadest components \citep[e.g.][]{2020ApJ...905...75H}, but this is far from universal.  For example, \cite{2024MNRAS.529..393N} identify an extremely variable AGN where the broadest double-peaked BLR component remained unchanged, while an intermediate width component changes rapidly.  This object (SDSS J125809.31+351943.0) appears qualitatively similar to the picture we see with \objname, where the broadest kinematic component is  approximately constant from 2001 to 2022, while the narrower component is only visible in the high-state spectra of 2015 and 2021.

The physical explanation given by \citet{2024MNRAS.529..393N} for J125809.31+351943.0 is that the broadest (constant) BLR component was within or near the disk, and heated by the hot X-ray emitting corona (see their Figure 11).  When the AGN changes to a low state [i.e.\ the low hard state, see \citet{2018MNRAS.480.3898N}], the hard ionising spectrum from the corona can be maintained.  In contrast, the narrower BLR seen in the high state can be ionised by the soft X-rays from the inner parts of the accretion disk.    If the accretion disk luminosity declines rapidly, then the high-state narrower BLR will also decline.  \objname\ shows similar characteristics to this picture.  In \objname\ the low-state BLR has higher obscuration, which is plausible if it is emitted in the outer disk, with a line of sight that might partially intersect a clumpy torus.

A useful test of the above picture would be evidence of a relatively hard X-ray spectrum in the low state.  \objname\ happens to be within the eROSITA Final Equatorial Depth Survey  \citep[eFEDS;][]{2022A&A...661A...1B} which was observed in November 2019, around the short term low state visible in the WISE photometry.  The flux of $\simeq1.7\times10^{-14}$\ergscm\ in the $0.2-2.3$\,keV band is not far above the detection limit for eROSITA.  We fit a simple power law to the 4 band eROSITA data to derive a photon index of $\Gamma=1.3^{+1.1}_{-0.8}$ (assuming a Galactic $N_H=3\times10^{20}$\,cm$^2$). While the value is relatively low compared to the typical $\Gamma\simeq1.8-2$ for AGN, the large uncertainties mean we cannot draw meaningful conclusions.  Deeper X-ray data would be required to examine this question in detail.

Finally, we note that the mean velocity of the BLR in the high state changes from $\simeq+170$ to $\simeq-190$\kms\ between 2015 and 2021 (relative to the rest frame defined by the galaxy's stellar population).  This could be driven by the motion of a clumpy disk wind across this period.  An obvious alternative is the presence of a binary black hole, such that we are seeing a periodic orbital velocity.  With current data it is hard to make a compelling case for this scenario, but further spectroscopy of the object when it next enters a high state would help to examine this picture.      

\section{Conclusion}\label{sec:conc}

We analyse spectra and photometry for galaxy \objname, which on the basis of SDSS and SAMI spectra is a changing-look AGN.  We draw the following conclusions:
\begin{itemize}
    \item \objname\ passes through at least one full changing-look cycle, as a Type 1.9 in 2001, a type 1 in 2015 and 2021, and a return to Type 1.9 in 2022.
    \item We find that the BLR changes in FWHM and $E(B-V)$ across the 21 years of observations.  The low state spectra (in 2001 and 2022) have larger FWHM $\simeq5000-6000$\,\kms\ and higher $E(B-V)\simeq6-8$\,mag.    The high state spectra (in 2015 and 2021) have lower FWHM$\simeq2000$\,\kms\ and lower $E(B-V)\simeq0.6-0.7$. 
    \item Based on measurements of the \dnfour\ index and independent polynomial fits, we find $\simeq10$ per cent of the spectral flux at 4000\AA\ is contributed by the AGN in the high state, while this is consistent with zero in the low state.  The high-state AGN continuum shape varies between 2015 and 2021, which could be due in part to flux calibration residuals, but in both cases they are redder than typical AGN continua, suggesting dust obscuration.
    \item Optical and infrared photometry show light curves consistent with the changes seen in the spectra, with by far the most significant change being in the WISE infrared bands.
    \item The WISE $W1-W2$ colours evolve from values typical of normal galaxies in the low state, to colours more typical of AGN in the high state.  The WISE photometry rules out the possibility that the changes we see are purely due to obscuration.  \objname\ is therefore a changing state AGN.
    \item A decline in WISE brightness in 2020 suggests that \objname\ may have briefly returned to a type 1.9 or type 2 state, before the rise to type 1 in 2021.
    \item The BLR structure of \objname\ is not consistent with a single component responding to a change in central luminosity.  Instead, we make the case that there are at least two separate components within the BLR. 
\end{itemize}

The number of changing-look AGN known has exploded in the last decade, driven by new time-domain imaging and spectroscopy.  A much smaller number have been shown, like \objname, to go through a complete cycle, but this number will no doubt rise in the coming decade.  The hope is that we can uncover the detailed structure of the region around the SMBH, which is still at a spatial resolution that is impossible to reach in the optical.

\section*{Acknowledgements}

The SAMI Galaxy Survey is based on observations made at the Anglo-Australian Telescope. The Sydney-AAO Multi-object Integral field spectrograph (SAMI) was developed jointly by the University of Sydney and the Australian Astronomical Observatory. The SAMI input catalogue is based on data taken from the Sloan Digital Sky Survey, the GAMA Survey and the VST ATLAS Survey. The SAMI Galaxy Survey is supported by the Australian Research Council Centre of Excellence for All Sky Astrophysics in 3 Dimensions (ASTRO 3D), through project number CE170100013, the Australian Research Council Centre of Excellence for All-sky Astrophysics (CAASTRO), through project number CE110001020, and other participating institutions. The SAMI Galaxy Survey website is http://sami-survey.org/ .

This publication makes use of data products from the Wide-field Infrared Survey Explorer, which is a joint project of the University of California, Los Angeles, and the Jet Propulsion Laboratory/California Institute of Technology, funded by the National Aeronautics and Space Administration.

This scientific work uses data obtained from Inyarrimanha Ilgari Bundara, the CSIRO Murchison Radio-astronomy Observatory. We acknowledge the Wajarri Yamaji People as the Traditional Owners and native title holders of the Observatory site. CSIRO’s ASKAP radio telescope is part of the Australia Telescope National Facility (https://ror.org/05qajvd42). Operation of ASKAP is funded by the Australian Government with support from the National Collaborative Research Infrastructure Strategy. ASKAP uses the resources of the Pawsey Supercomputing Research Centre. Establishment of ASKAP, Inyarrimanha Ilgari Bundara, the CSIRO Murchison Radio-astronomy Observatory and the Pawsey Supercomputing Research Centre are initiatives of the Australian Government, with support from the Government of Western Australia and the Science and Industry Endowment Fund.

Funding for the SDSS and SDSS-II has been provided by the Alfred P. Sloan Foundation, the Participating Institutions, the National Science Foundation, the U.S. Department of Energy, the National Aeronautics and Space Administration, the Japanese Monbukagakusho, the Max Planck Society, and the Higher Education Funding Council for England. The SDSS Web Site is http://www.sdss.org/.

The SDSS is managed by the Astrophysical Research Consortium for the Participating Institutions. The Participating Institutions are the American Museum of Natural History, Astrophysical Institute Potsdam, University of Basel, University of Cambridge, Case Western Reserve University, University of Chicago, Drexel University, Fermilab, the Institute for Advanced Study, the Japan Participation Group, Johns Hopkins University, the Joint Institute for Nuclear Astrophysics, the Kavli Institute for Particle Astrophysics and Cosmology, the Korean Scientist Group, the Chinese Academy of Sciences (LAMOST), Los Alamos National Laboratory, the Max-Planck-Institute for Astronomy (MPIA), the Max-Planck-Institute for Astrophysics (MPA), New Mexico State University, Ohio State University, University of Pittsburgh, University of Portsmouth, Princeton University, the United States Naval Observatory, and the University of Washington. 

OT and SP acknowledge funding Denison Scholarship funding from the Faculty of Science, University of Sydney.

This paper includes archived data obtained through the CSIRO ASKAP Science Data Archive, CASDA (http://data.csiro.au).

\bibliographystyle{paslike}
\bibliography{references}

@ARTICLE{1995ApJ...455L.119B,
       author = {{Baldwin}, Jack and {Ferland}, Gary and {Korista}, Kirk and {Verner}, Dima},
        title = "{Locally Optimally Emitting Clouds and the Origin of Quasar Emission Lines}",
      journal = {ApJL},
         year = 1995,
        month = dec,
       volume = {455},
        pages = {L119},
          doi = {10.1086/309827},
archivePrefix = {arXiv},
       eprint = {astro-ph/9510080},
 primaryClass = {astro-ph},
       adsurl = {https://ui.adsabs.harvard.edu/abs/1995ApJ...455L.119B}
}

@article{RevModPhys.51.715,
  title = {The emission lines of quasars and similar objects},
  author = {Davidson, Kris and Netzer, Hagai},
  journal = {Rev. Mod. Phys.},
  volume = {51},
  issue = {4},
  pages = {715--766},
  numpages = {0},
  year = {1979},
  month = {Oct},
  publisher = {American Physical Society},
  doi = {10.1103/RevModPhys.51.715},
  url = {https://link.aps.org/doi/10.1103/RevModPhys.51.715}
}

@ARTICLE{2016MNRAS.462.3570S,
       author = {{Schnorr-M{\"u}ller}, Allan and {Davies}, R.~I. and {Korista}, K.~T. and {Burtscher}, L. and {Rosario}, D. and {Storchi-Bergmann}, T. and {Contursi}, A. and {Genzel}, R. and {Graci{\'a}-Carpio}, J. and {Hicks}, E.~K.~S. and {Janssen}, A. and {Koss}, M. and {Lin}, M.-Y. and {Lutz}, D. and {Maciejewski}, W. and {M{\"u}ller-S{\'a}nchez}, F. and {Orban de Xivry}, G. and {Riffel}, R. and {Riffel}, Rogemar A. and {Schartmann}, M. and {Sternberg}, A. and {Sturm}, E. and {Tacconi}, L. and {Veilleux}, S. and {Ulrich}, O.~A.},
        title = "{Constraints on the broad-line region properties and extinction in local Seyferts}",
      journal = {\mnras},
         year = 2016,
        month = nov,
       volume = {462},
       number = {4},
        pages = {3570-3590},
          doi = {10.1093/mnras/stw1865},
archivePrefix = {arXiv},
       eprint = {1607.07308},
 primaryClass = {astro-ph.GA},
       adsurl = {https://ui.adsabs.harvard.edu/abs/2016MNRAS.462.3570S}
}

@ARTICLE{2022MNRAS.509.4940T,
       author = {{Thorne}, Jessica E. and {Robotham}, Aaron S.~G. and {Davies}, Luke J.~M. and {Bellstedt}, Sabine and {Brown}, Michael J.~I. and {Croom}, Scott M. and {Delvecchio}, Ivan and {Groves}, Brent and {Jarvis}, Matt J. and {Shabala}, Stanislav S. and {Seymour}, Nick and {Whittam}, Imogen H. and {Bravo}, Matias and {Cook}, Robin H.~W. and {Driver}, Simon P. and {Holwerda}, Benne and {Phillipps}, Steven and {Siudek}, Malgorzata},
        title = "{Deep Extragalactic VIsible Legacy Survey (DEVILS): identification of AGN through SED fitting and the evolution of the bolometric AGN luminosity function}",
      journal = {MNRAS},
         year = 2022,
        month = feb,
       volume = {509},
       number = {4},
        pages = {4940-4961},
          doi = {10.1093/mnras/stab3208},
archivePrefix = {arXiv},
       eprint = {2112.06366},
 primaryClass = {astro-ph.GA},
       adsurl = {https://ui.adsabs.harvard.edu/abs/2022MNRAS.509.4940T}
}

@ARTICLE{2017ApJ...850...68C,
       author = {{Cluver}, M.~E. and {Jarrett}, T.~H. and {Dale}, D.~A. and {Smith}, J.-D.~T. and {August}, Tamlyn and {Brown}, M.~J.~I.},
        title = "{Calibrating Star Formation in WISE Using Total Infrared Luminosity}",
      journal = {\apj},
         year = 2017,
        month = nov,
       volume = {850},
       number = {1},
          eid = {68},
        pages = {68},
          doi = {10.3847/1538-4357/aa92c7},
archivePrefix = {arXiv},
       eprint = {1710.03469},
 primaryClass = {astro-ph.GA},
       adsurl = {https://ui.adsabs.harvard.edu/abs/2017ApJ...850...68C}
}

@ARTICLE{2026arXiv260715564Z,
       author = {{Zhu}, Litao and {Wang}, Zhongxiang and {Gupta}, Alok C. and {Zeng}, Yiyang and {Devanand}, P.~U. and {Dogra}, Karan and {Bhowmick}, Riya and {Zhang}, Ju-Jia},
        title = "{22 Newly Identified Repeating Changing-look AGNs and Evidence for Extreme Broad-line Region Breathing}",
      journal = {arXiv},
         year = 2026,
        month = jul,
          eid = {arXiv:2607.15564},
        pages = {arXiv:2607.15564},
          doi = {10.48550/arXiv.2607.15564},
archivePrefix = {arXiv},
       eprint = {2607.15564},
 primaryClass = {astro-ph.GA},
       adsurl = {https://ui.adsabs.harvard.edu/abs/2026arXiv260715564Z}
}

@ARTICLE{2022MNRAS.511...54H,
       author = {{Hon}, Wei Jeat and {Wolf}, Christian and {Onken}, Christopher A. and {Webster}, Rachel and {Auchettl}, Katie},
        title = "{SkyMapper colours of Seyfert galaxies and changing-look AGN - II. Newly discovered changing-look AGN}",
      journal = {MNRAS},
         year = 2022,
        month = mar,
       volume = {511},
       number = {1},
        pages = {54-70},
          doi = {10.1093/mnras/stab3694},
       adsurl = {https://ui.adsabs.harvard.edu/abs/2022MNRAS.511...54H}
}

@ARTICLE{2022AJ....164..207D,
       author = {{DESI Collaboration} and {Abareshi}, B. and {Aguilar}, J. and {Ahlen}, S. and {Alam}, Shadab and {Alexander}, David M. and {Alfarsy}, R. and {Allen}, L. and {Allende Prieto}, C. and {Alves}, O. and {Ameel}, J. and {Armengaud}, E. and {Asorey}, J. and {Aviles}, Alejandro and {Bailey}, S. and {Balaguera-Antol{\'\i}nez}, A. and {Ballester}, O. and {Baltay}, C. and {Bault}, A. and {Beltran}, S.~F. and {Benavides}, B. and {BenZvi}, S. and {Berti}, A. and {Besuner}, R. and {Beutler}, Florian and {Bianchi}, D. and {Blake}, C. and {Blanc}, P. and {Blum}, R. and {Bolton}, A. and {Bose}, S. and {Bramall}, D. and {Brieden}, S. and {Brodzeller}, A. and {Brooks}, D. and {Brownewell}, C. and {Buckley-Geer}, E. and {Cahn}, R.~N. and {Cai}, Z. and {Canning}, R. and {Capasso}, R. and {Carnero Rosell}, A. and {Carton}, P. and {Casas}, R. and {Castander}, F.~J. and {Cervantes-Cota}, J.~L. and {Chabanier}, S. and {Chaussidon}, E. and {Chuang}, C. and {Circosta}, C. and {Cole}, S. and {Cooper}, A.~P. and {da Costa}, L. and {Cousinou}, M.-C. and {Cuceu}, A. and {Davis}, T.~M. and {Dawson}, K. and {de la Cruz-Noriega}, R. and {de la Macorra}, A. and {de Mattia}, A. and {Della Costa}, J. and {Demmer}, P. and {Derwent}, M. and {Dey}, A. and {Dey}, B. and {Dhungana}, G. and {Ding}, Z. and {Dobson}, C. and {Doel}, P. and {Donald-McCann}, J. and {Donaldson}, J. and {Douglass}, K. and {Duan}, Y. and {Dunlop}, P. and {Edelstein}, J. and {Eftekharzadeh}, S. and {Eisenstein}, D.~J. and {Enriquez-Vargas}, M. and {Escoffier}, S. and {Evatt}, M. and {Fagrelius}, P. and {Fan}, X. and {Fanning}, K. and {Fawcett}, V.~A. and {Ferraro}, S. and {Ereza}, J. and {Flaugher}, B. and {Font-Ribera}, A. and {Forero-Romero}, J.~E. and {Frenk}, C.~S. and {Fromenteau}, S. and {G{\"a}nsicke}, B.~T. and {Garcia-Quintero}, C. and {Garrison}, L. and {Gazta{\~n}aga}, E. and {Gerardi}, F. and {Gil-Mar{\'\i}n}, H. and {Gontcho A Gontcho}, S. and {Gonzalez-Morales}, Alma X. and {Gonzalez-de-Rivera}, G. and {Gonzalez-Perez}, V. and {Gordon}, C. and {Graur}, O. and {Green}, D. and {Grove}, C. and {Gruen}, D. and {Gutierrez}, G. and {Guy}, J. and {Hahn}, C. and {Harris}, S. and {Herrera}, D. and {Herrera-Alcantar}, Hiram K. and {Honscheid}, K. and {Howlett}, C. and {Huterer}, D. and {Ir{\v{s}}i{\v{c}}}, V. and {Ishak}, M. and {Jelinsky}, P. and {Jiang}, L. and {Jimenez}, J. and {Jing}, Y.~P. and {Joyce}, R. and {Jullo}, E. and {Juneau}, S. and {Kara{\c{c}}ayl{\i}}, N.~G. and {Karamanis}, M. and {Karcher}, A. and {Karim}, T. and {Kehoe}, R. and {Kent}, S. and {Kirkby}, D. and {Kisner}, T. and {Kitaura}, F. and {Koposov}, S.~E. and {Kov{\'a}cs}, A. and {Kremin}, A. and {Krolewski}, Alex and {L'Huillier}, B. and {Lahav}, O. and {Lambert}, A. and {Lamman}, C. and {Lan}, Ting-Wen and {Landriau}, M. and {Lane}, S. and {Lang}, D. and {Lange}, J.~U. and {Lasker}, J. and {Le Guillou}, L. and {Leauthaud}, A. and {Le Van Suu}, A. and {Levi}, Michael E. and {Li}, T.~S. and {Magneville}, C. and {Manera}, M. and {Manser}, Christopher J. and {Marshall}, B. and {Martini}, Paul and {McCollam}, W. and {McDonald}, P. and {Meisner}, Aaron M. and {Mena-Fern{\'a}ndez}, J. and {Meneses-Rizo}, J. and {Mezcua}, M. and {Miller}, T. and {Miquel}, R. and {Montero-Camacho}, P. and {Moon}, J. and {Moustakas}, J. and {Mueller}, E. and {Mu{\~n}oz-Guti{\'e}rrez}, Andrea and {Myers}, Adam D. and {Nadathur}, S. and {Najita}, J. and {Napolitano}, L. and {Neilsen}, E. and {Newman}, Jeffrey A. and {Nie}, J.~D. and {Ning}, Y. and {Niz}, G. and {Norberg}, P. and {Noriega}, Hern{\'a}n E. and {O'Brien}, T. and {Obuljen}, A. and {Palanque-Delabrouille}, N. and {Palmese}, A. and {Zhiwei}, P. and {Pappalardo}, D. and {PENG}, X. and {Percival}, W.~J. and {Perruchot}, S. and {Pogge}, R. and {Poppett}, C. and {Porredon}, A. and {Prada}, F. and {Prochaska}, J. and {Pucha}, R. and {P{\'e}rez-Fern{\'a}ndez}, A. and {P{\'e}rez-R{\`a}fols}, I. and {Rabinowitz}, D. and {Raichoor}, A.},
        title = "{Overview of the Instrumentation for the Dark Energy Spectroscopic Instrument}",
      journal = {AJ},
         year = 2022,
        month = nov,
       volume = {164},
       number = {5},
          eid = {207},
        pages = {207},
          doi = {10.3847/1538-3881/ac882b},
archivePrefix = {arXiv},
       eprint = {2205.10939},
 primaryClass = {astro-ph.IM},
       adsurl = {https://ui.adsabs.harvard.edu/abs/2022AJ....164..207D}
}

@ARTICLE{1976ApJ...210L.117T,
       author = {{Tohline}, J.~E. and {Osterbrock}, D.~E.},
        title = "{Variation of the spectrum of the Seyfert galaxy NGC 7603.}",
      journal = {ApJL},
         year = 1976,
        month = dec,
       volume = {210},
        pages = {L117-L120},
          doi = {10.1086/182317},
       adsurl = {https://ui.adsabs.harvard.edu/abs/1976ApJ...210L.117T}
}

@ARTICLE{1971ApJ...164L.109K,
       author = {{Khachikian}, Edward Ye. and {Weedman}, Daniel W.},
        title = "{A New Cloud of Hydrogen Emission in a Bright Galactic Nucleus}",
      journal = {ApJL},
         year = 1971,
        month = mar,
       volume = {164},
        pages = {L109},
          doi = {10.1086/180701},
       adsurl = {https://ui.adsabs.harvard.edu/abs/1971ApJ...164L.109K}
}

@ARTICLE{2024MNRAS.535.2322A,
       author = {{Amrutha}, Neelesh and {Wolf}, Christian and {Onken}, Christopher A. and {Hon}, Wei Jeat and {Lai}, Samuel and {Tonry}, John L. and {Webster}, Rachel},
        title = "{Discovering changing-look AGN in the 6dF Galaxy Survey using ATLAS light curves}",
      journal = {MNRAS},
         year = 2024,
        month = dec,
       volume = {535},
       number = {3},
        pages = {2322-2340},
          doi = {10.1093/mnras/stae2470},
archivePrefix = {arXiv},
       eprint = {2410.22671},
 primaryClass = {astro-ph.GA},
       adsurl = {https://ui.adsabs.harvard.edu/abs/2024MNRAS.535.2322A}
}

@ARTICLE{1981PASP...93....5B,
       author = {{Baldwin}, J.~A. and {Phillips}, M.~M. and {Terlevich}, R.},
        title = "{Classification parameters for the emission-line spectra of extragalactic objects.}",
      journal = {PASP},
         year = 1981,
        month = feb,
       volume = {93},
        pages = {5-19},
          doi = {10.1086/130766},
       adsurl = {https://ui.adsabs.harvard.edu/abs/1981PASP...93....5B}
}

@ARTICLE{2023MNRAS.518.2177G,
       author = {{Graham}, Alister W. and {Sahu}, Nandini},
        title = "{Appreciating mergers for understanding the non-linear M$_{bh}$-M$_{*,spheroid}$ and M$_{bh}$-M$_{*, galaxy}$ relations, updated herein, and the implications for the (reduced) role of AGN feedback}",
      journal = {MNRAS},
         year = 2023,
        month = jan,
       volume = {518},
       number = {2},
        pages = {2177-2200},
          doi = {10.1093/mnras/stac2019},
archivePrefix = {arXiv},
       eprint = {2209.14526},
 primaryClass = {astro-ph.GA},
       adsurl = {https://ui.adsabs.harvard.edu/abs/2023MNRAS.518.2177G}
}

@ARTICLE{2001ApJS..134....1V,
       author = {{Vestergaard}, M. and {Wilkes}, B.~J.},
        title = "{An Empirical Ultraviolet Template for Iron Emission in Quasars as Derived from I Zwicky 1}",
      journal = {ApJS},
         year = 2001,
        month = may,
       volume = {134},
       number = {1},
        pages = {1-33},
          doi = {10.1086/320357},
archivePrefix = {arXiv},
       eprint = {astro-ph/0104320},
 primaryClass = {astro-ph},
       adsurl = {https://ui.adsabs.harvard.edu/abs/2001ApJS..134....1V}
}

@article{DAVIDSON2008162,
title = {The wild bootstrap, tamed at last},
journal = {Journal of Econometrics},
volume = {146},
number = {1},
pages = {162-169},
year = {2008},
issn = {0304-4076},
doi = {https://doi.org/10.1016/j.jeconom.2008.08.003},
url = {https://www.sciencedirect.com/science/article/pii/S0304407608000833},
author = {Russell Davidson and Emmanuel Flachaire}
}

@ARTICLE{2016ApJ...831..157M,
       author = {{Margala}, Daniel and {Kirkby}, David and {Dawson}, Kyle and {Bailey}, Stephen and {Blanton}, Michael and {Schneider}, Donald P.},
        title = "{Improved Spectrophotometric Calibration of the SDSS-III BOSS Quasar Sample}",
      journal = {ApJ},
         year = 2016,
        month = nov,
       volume = {831},
       number = {2},
          eid = {157},
        pages = {157},
          doi = {10.3847/0004-637X/831/2/157},
archivePrefix = {arXiv},
       eprint = {1506.04790},
 primaryClass = {astro-ph.IM},
       adsurl = {https://ui.adsabs.harvard.edu/abs/2016ApJ...831..157M}
}

@ARTICLE{2022A&A...661A...1B,
       author = {{Brunner}, H. and {Liu}, T. and {Lamer}, G. and {Georgakakis}, A. and {Merloni}, A. and {Brusa}, M. and {Bulbul}, E. and {Dennerl}, K. and {Friedrich}, S. and {Liu}, A. and {Maitra}, C. and {Nandra}, K. and {Ramos-Ceja}, M.~E. and {Sanders}, J.~S. and {Stewart}, I.~M. and {Boller}, T. and {Buchner}, J. and {Clerc}, N. and {Comparat}, J. and {Dwelly}, T. and {Eckert}, D. and {Finoguenov}, A. and {Freyberg}, M. and {Ghirardini}, V. and {Gueguen}, A. and {Haberl}, F. and {Kreykenbohm}, I. and {Krumpe}, M. and {Osterhage}, S. and {Pacaud}, F. and {Predehl}, P. and {Reiprich}, T.~H. and {Robrade}, J. and {Salvato}, M. and {Santangelo}, A. and {Schrabback}, T. and {Schwope}, A. and {Wilms}, J.},
        title = "{The eROSITA Final Equatorial Depth Survey (eFEDS). X-ray catalogue}",
      journal = {A\&A},
         year = 2022,
        month = may,
       volume = {661},
          eid = {A1},
        pages = {A1},
          doi = {10.1051/0004-6361/202141266},
archivePrefix = {arXiv},
       eprint = {2106.14517},
 primaryClass = {astro-ph.HE},
       adsurl = {https://ui.adsabs.harvard.edu/abs/2022A&A...661A...1B}
}

@ARTICLE{1981ApJ...249..462O,
       author = {{Osterbrock}, D.~E.},
        title = "{Seyfert galaxies with weak broad H alpha emission lines}",
      journal = {\apj},
         year = 1981,
        month = oct,
       volume = {249},
        pages = {462-470},
          doi = {10.1086/159306},
       adsurl = {https://ui.adsabs.harvard.edu/abs/1981ApJ...249..462O}
}

@ARTICLE{2025ApJS..278...28G,
       author = {{Guo}, Wei-Jian and {Zou}, Hu and {Greenwell}, Claire L. and {Alexander}, David M. and {Fawcett}, Victoria A. and {Pan}, Zhiwei and {Siudek}, Ma{\l}gorzata and {Aguilar}, Jessica Nicole and {Ahlen}, Steven and {Brooks}, David and {Claybaugh}, Todd and {Dawson}, Kyle and {de la Macorra}, Axel and {Doel}, Peter and {Font-Ribera}, Andreu and {Gazta{\~n}aga}, Enrique and {Gontcho A Gontcho}, Satya and {Gutierrez}, Gaston and {Kehoe}, Robert and {Kisner}, Theodore and {Landriau}, Martin and {Le Guillou}, Laurent and {Manera}, Marc and {Meisner}, Aaron and {Miquel}, Ramon and {Moustakas}, John and {Prada}, Francisco and {Rossi}, Graziano and {Sanchez}, Eusebio and {Schubnell}, Michael and {Sprayberry}, David and {Sui}, Jipeng and {Tarl{\'e}}, Gregory and {Weaver}, Benjamin Alan and {Xiao}, Yun-Ao and {Zou}, Siwei},
        title = "{Changing-look Active Galactic Nuclei from the Dark Energy Spectroscopic Instrument. II. Statistical Properties from the First Data Release}",
      journal = {\apjs},
         year = 2025,
        month = may,
       volume = {278},
       number = {1},
          eid = {28},
        pages = {28},
          doi = {10.3847/1538-4365/adc124},
archivePrefix = {arXiv},
       eprint = {2408.00402},
 primaryClass = {astro-ph.GA},
       adsurl = {https://ui.adsabs.harvard.edu/abs/2025ApJS..278...28G}
}

@ARTICLE{2024MNRAS.529..393N,
       author = {{Nagoshi}, Shumpei and {Iwamuro}, Fumihide and {Yamada}, Satoshi and {Ueda}, Yoshihiro and {Oikawa}, Yuto and {Otsuka}, Masaaki and {Isogai}, Keisuke and {Mineshige}, Shin},
        title = "{Probing the origin of the two-component structure of broad-line region by reverberation mapping of an extremely variable quasar}",
      journal = {MNRAS},
         year = 2024,
        month = mar,
       volume = {529},
       number = {1},
        pages = {393-408},
          doi = {10.1093/mnras/stae319},
archivePrefix = {arXiv},
       eprint = {2306.13930},
 primaryClass = {astro-ph.GA},
       adsurl = {https://ui.adsabs.harvard.edu/abs/2024MNRAS.529..393N}
}

@ARTICLE{2018MNRAS.480.3898N,
       author = {{Noda}, Hirofumi and {Done}, Chris},
        title = "{Explaining changing-look AGN with state transition triggered by rapid mass accretion rate drop}",
      journal = {\mnras},
         year = 2018,
        month = nov,
       volume = {480},
       number = {3},
        pages = {3898-3906},
          doi = {10.1093/mnras/sty2032},
archivePrefix = {arXiv},
       eprint = {1805.07873},
 primaryClass = {astro-ph.GA},
       adsurl = {https://ui.adsabs.harvard.edu/abs/2018MNRAS.480.3898N}
}

@ARTICLE{2022ApJ...925...52U,
       author = {{U}, Vivian and {Barth}, Aaron J. and {Vogler}, H. Alexander and {Guo}, Hengxiao and {Treu}, Tommaso and {Bennert}, Vardha N. and {Canalizo}, Gabriela and {Filippenko}, Alexei V. and {Gates}, Elinor and {Hamann}, Frederick and {Joner}, Michael D. and {Malkan}, Matthew A. and {Pancoast}, Anna and {Williams}, Peter R. and {Woo}, Jong-Hak and {Abolfathi}, Bela and {Abramson}, L.~E. and {Armen}, Stephen F. and {Bae}, Hyun-Jin and {Bohn}, Thomas and {Boizelle}, Benjamin D. and {Bostroem}, Azalee and {Brandel}, Andrew and {Brink}, Thomas G. and {Channa}, Sanyum and {Cooper}, M.~C. and {Cosens}, Maren and {Donohue}, Edward and {Fillingham}, Sean P. and {Gonz{\'a}lez-Buitrago}, Diego and {Halevi}, Goni and {Halle}, Andrew and {Hood}, Carol E. and {Horne}, Keith and {Horst}, J. Chuck and {de Kouchkovsky}, Maxime and {Kuhn}, Benjamin and {Kumar}, Sahana and {Leonard}, Douglas C. and {Loveland}, Donald and {Manzano-King}, Christina and {McHardy}, Ian and {Michel}, Ra{\'u}l and {Olaes}, Melanie Kae B. and {Park}, Daeseong and {Park}, Songyoun and {Pei}, Liuyi and {Ross}, Timothy W. and {Runco}, Jordan N. and {Samuel}, Jenna and {S{\'a}nchez}, Javier and {Scott}, Bryan and {Sexton}, Remington O. and {Shin}, Jaejin and {Shivvers}, Isaac and {Spencer}, Chance L. and {Stahl}, Benjamin E. and {Stegman}, Samantha and {Stomberg}, Isak and {Valenti}, Stefano and {Villafa{\~n}a}, L. and {Walsh}, Jonelle L. and {Yuk}, Heechan and {Zheng}, WeiKang},
        title = "{The Lick AGN Monitoring Project 2016: Velocity-resolved H{\ensuremath{\beta}} Lags in Luminous Seyfert Galaxies}",
      journal = {\apj},
         year = 2022,
        month = jan,
       volume = {925},
       number = {1},
          eid = {52},
        pages = {52},
          doi = {10.3847/1538-4357/ac3d26},
archivePrefix = {arXiv},
       eprint = {2111.14849},
 primaryClass = {astro-ph.GA},
       adsurl = {https://ui.adsabs.harvard.edu/abs/2022ApJ...925...52U}
}

@ARTICLE{2016ApJ...820...27D,
       author = {{Du}, Pu and {Lu}, Kai-Xing and {Hu}, Chen and {Qiu}, Jie and {Li}, Yan-Rong and {Huang}, Ying-Ke and {Wang}, Fang and {Bai}, Jin-Ming and {Bian}, Wei-Hao and {Yuan}, Ye-Fei and {Ho}, Luis C. and {Wang}, Jian-Min and {SEAMBH Collaboration}},
        title = "{Supermassive Black Holes with High Accretion Rates in Active Galactic Nuclei. VI. Velocity-resolved Reverberation Mapping of the H{\ensuremath{\beta}} Line}",
      journal = {\apj},
         year = 2016,
        month = mar,
       volume = {820},
       number = {1},
          eid = {27},
        pages = {27},
          doi = {10.3847/0004-637X/820/1/27},
archivePrefix = {arXiv},
       eprint = {1602.01922},
 primaryClass = {astro-ph.GA},
       adsurl = {https://ui.adsabs.harvard.edu/abs/2016ApJ...820...27D}
}

@ARTICLE{2003A&A...407..461K,
       author = {{Kollatschny}, W.},
        title = "{Accretion disk wind in the AGN broad-line region: Spectroscopically resolved line profile variations in Mrk 110}",
      journal = {A\&A},
         year = 2003,
        month = aug,
       volume = {407},
        pages = {461-472},
          doi = {10.1051/0004-6361:20030928},
archivePrefix = {arXiv},
       eprint = {astro-ph/0306389},
 primaryClass = {astro-ph},
       adsurl = {https://ui.adsabs.harvard.edu/abs/2003A&A...407..461K}
}

@ARTICLE{2020ApJ...905...75H,
       author = {{Hu}, Chen and {Li}, Sha-Sha and {Guo}, Wei-Jian and {Yang}, Sen and {Yang}, Zi-Xu and {Bao}, Dong-Wei and {Jiang}, Bo-Wei and {Du}, Pu and {Li}, Yan-Rong and {Xiao}, Ming and {Songsheng}, Yu-Yang and {Yu}, Zhe and {Bai}, Jin-Ming and {Ho}, Luis C. and {Bian}, Wei-Hao and {Brotherton}, Michael S. and {Yuan}, Ye-Fei and {Aceituno}, Jes{\'u}s and {Winkler}, Hartmut and {Wang}, Jian-Min and {SEAMBH Collaboration}},
        title = "{Evidence for Two Distinct Broad-line Regions from Reverberation Mapping of PG 0026+129}",
      journal = {ApJ},
         year = 2020,
        month = dec,
       volume = {905},
       number = {1},
          eid = {75},
        pages = {75},
          doi = {10.3847/1538-4357/abc2da},
archivePrefix = {arXiv},
       eprint = {2010.09871},
 primaryClass = {astro-ph.GA},
       adsurl = {https://ui.adsabs.harvard.edu/abs/2020ApJ...905...75H}
}

@ARTICLE{2018ApJ...866..123M,
       author = {{Mathur}, S. and {Denney}, K.~D. and {Gupta}, A. and {Vestergaard}, M. and {De Rosa}, G. and {Krongold}, Yair and {Nicastro}, F. and {Collinson}, J. and {Goad}, M. and {Korista}, K. and {Pogge}, R.~W. and {Peterson}, B.~M.},
        title = "{The Changing-look Quasar Mrk 590 Is Awakening}",
      journal = {\apj},
         year = 2018,
        month = oct,
       volume = {866},
       number = {2},
          eid = {123},
        pages = {123},
          doi = {10.3847/1538-4357/aadd91},
archivePrefix = {arXiv},
       eprint = {1810.06616},
 primaryClass = {astro-ph.GA},
       adsurl = {https://ui.adsabs.harvard.edu/abs/2018ApJ...866..123M}
}

@ARTICLE{2014ApJ...796..134D,
       author = {{Denney}, K.~D. and {De Rosa}, G. and {Croxall}, K. and {Gupta}, A. and {Bentz}, M.~C. and {Fausnaugh}, M.~M. and {Grier}, C.~J. and {Martini}, P. and {Mathur}, S. and {Peterson}, B.~M. and {Pogge}, R.~W. and {Shappee}, B.~J.},
        title = "{The Typecasting of Active Galactic Nuclei: Mrk 590 no Longer Fits the Role}",
      journal = {\apj},
         year = 2014,
        month = dec,
       volume = {796},
       number = {2},
          eid = {134},
        pages = {134},
          doi = {10.1088/0004-637X/796/2/134},
archivePrefix = {arXiv},
       eprint = {1404.4879},
 primaryClass = {astro-ph.GA},
       adsurl = {https://ui.adsabs.harvard.edu/abs/2014ApJ...796..134D}
}

@ARTICLE{1986ApJ...311..135C,
       author = {{Cohen}, Ross D. and {Rudy}, Richard J. and {Puetter}, R.~C. and {Ake}, T.~B. and {Foltz}, Craig B.},
        title = "{Variability of Markarian 1018: Seyfert 1.9 to Seyfert 1}",
      journal = {\apj},
         year = 1986,
        month = dec,
       volume = {311},
        pages = {135},
          doi = {10.1086/164758},
       adsurl = {https://ui.adsabs.harvard.edu/abs/1986ApJ...311..135C}
}

@ARTICLE{2025ApJ...981..129W,
       author = {{Wang}, Shu and {Woo}, Jong-Hak and {Gallo}, Elena and {Son}, Donghoon and {Yang}, Qian and {Jin}, Junjie and {Guo}, Hengxiao and {Kong}, Minzhi},
        title = "{Dormancy and Reawakening over Years: Eight New Recurrent Changing-look AGNs}",
      journal = {ApJ},
         year = 2025,
        month = mar,
       volume = {981},
       number = {2},
          eid = {129},
        pages = {129},
          doi = {10.3847/1538-4357/adadf3},
archivePrefix = {arXiv},
       eprint = {2410.15587},
 primaryClass = {astro-ph.GA},
       adsurl = {https://ui.adsabs.harvard.edu/abs/2025ApJ...981..129W}
}

@ARTICLE{2025ApJ...986..160D,
       author = {{Dong}, Qian and {Zhang}, Zhi-Xiang and {Gu}, Wei-Min and {Sun}, Mouyuan and {Zheng}, Yong-Gang},
        title = "{Newly Discovered Changing-look Active Galactic Nuclei from SDSS and LAMOST Survey}",
      journal = {ApJ},
         year = 2025,
        month = jun,
       volume = {986},
       number = {2},
          eid = {160},
        pages = {160},
          doi = {10.3847/1538-4357/add331},
archivePrefix = {arXiv},
       eprint = {2408.07335},
 primaryClass = {astro-ph.GA},
       adsurl = {https://ui.adsabs.harvard.edu/abs/2025ApJ...986..160D}
}

@ARTICLE{2025arXiv251018445D,
       author = {{Dong}, Qian and {Zhang}, Zhi-Xiang and {Gu}, Wei-Min and {Sun}, Mouyuan and {Guo}, Wei-Jian and {Cai}, Zhen-Yi and {Wang}, Jun-Xian and {Zheng}, Yong-Gang},
        title = "{Discovery of Repeating Transitions in 25 Changing-look Active Galactic Nuclei}",
      journal = {arXiv},
         year = 2025,
        month = oct,
          eid = {arXiv:2510.18445},
        pages = {arXiv:2510.18445},
          doi = {10.48550/arXiv.2510.18445},
archivePrefix = {arXiv},
       eprint = {2510.18445},
 primaryClass = {astro-ph.GA},
       adsurl = {https://ui.adsabs.harvard.edu/abs/2025arXiv251018445D}
}

@ARTICLE{1992ARA&A..30..575C,
       author = {{Condon}, J.~J.},
        title = "{Radio emission from normal galaxies.}",
      journal = {ARA\&A},
         year = 1992,
        month = jan,
       volume = {30},
        pages = {575-611},
          doi = {10.1146/annurev.aa.30.090192.003043},
       adsurl = {https://ui.adsabs.harvard.edu/abs/1992ARA&A..30..575C}
}

@ARTICLE{2011ApJS..194...29R,
       author = {{Richards}, Joseph L. and {Max-Moerbeck}, Walter and {Pavlidou}, Vasiliki and {King}, Oliver G. and {Pearson}, Timothy J. and {Readhead}, Anthony C.~S. and {Reeves}, Rodrigo and {Shepherd}, Martin C. and {Stevenson}, Matthew A. and {Weintraub}, Lawrence C. and {Fuhrmann}, Lars and {Angelakis}, Emmanouil and {Zensus}, J. Anton and {Healey}, Stephen E. and {Romani}, Roger W. and {Shaw}, Michael S. and {Grainge}, Keith and {Birkinshaw}, Mark and {Lancaster}, Katy and {Worrall}, Diana M. and {Taylor}, Gregory B. and {Cotter}, Garret and {Bustos}, Ricardo},
        title = "{Blazars in the Fermi Era: The OVRO 40 m Telescope Monitoring Program}",
      journal = {ApJS},
         year = 2011,
        month = jun,
       volume = {194},
       number = {2},
          eid = {29},
        pages = {29},
          doi = {10.1088/0067-0049/194/2/29},
archivePrefix = {arXiv},
       eprint = {1011.3111},
 primaryClass = {astro-ph.CO},
       adsurl = {https://ui.adsabs.harvard.edu/abs/2011ApJS..194...29R}
}

@ARTICLE{1962SvA.....6..317K,
       author = {{Kardashev}, N.~S.},
        title = "{Nonstationarity of Spectra of Young Sources of Nonthermal Radio Emission}",
      journal = {SvA},
         year = 1962,
        month = dec,
       volume = {6},
        pages = {317},
       adsurl = {https://ui.adsabs.harvard.edu/abs/1962SvA.....6..317K}
}

@ARTICLE{2000AJ....119.1111B,
       author = {{Blundell}, Katherine M. and {Rawlings}, Steve},
        title = "{The Spectra and Energies of Classical Double Radio Lobes}",
      journal = {AJ},
         year = 2000,
        month = mar,
       volume = {119},
       number = {3},
        pages = {1111-1122},
          doi = {10.1086/301254},
archivePrefix = {arXiv},
       eprint = {astro-ph/0001327},
 primaryClass = {astro-ph},
       adsurl = {https://ui.adsabs.harvard.edu/abs/2000AJ....119.1111B}
}

@ARTICLE{2017MNRAS.466.2888H,
       author = {{Harwood}, Jeremy J.},
        title = "{Spectral ageing in the era of big data: integrated versus resolved models}",
      journal = {MNRAS},
         year = 2017,
        month = apr,
       volume = {466},
       number = {3},
        pages = {2888-2894},
          doi = {10.1093/mnras/stw3318},
archivePrefix = {arXiv},
       eprint = {1612.04390},
 primaryClass = {astro-ph.GA},
       adsurl = {https://ui.adsabs.harvard.edu/abs/2017MNRAS.466.2888H}
}

@ARTICLE{2011MNRAS.413..971D,
       author = {{Driver}, S.~P. and {Hill}, D.~T. and {Kelvin}, L.~S. and {Robotham}, A.~S.~G. and {Liske}, J. and {Norberg}, P. and {Baldry}, I.~K. and {Bamford}, S.~P. and {Hopkins}, A.~M. and {Loveday}, J. and {Peacock}, J.~A. and {Andrae}, E. and {Bland-Hawthorn}, J. and {Brough}, S. and {Brown}, M.~J.~I. and {Cameron}, E. and {Ching}, J.~H.~Y. and {Colless}, M. and {Conselice}, C.~J. and {Croom}, S.~M. and {Cross}, N.~J.~G. and {de Propris}, R. and {Dye}, S. and {Drinkwater}, M.~J. and {Ellis}, S. and {Graham}, Alister W. and {Grootes}, M.~W. and {Gunawardhana}, M. and {Jones}, D.~H. and {van Kampen}, E. and {Maraston}, C. and {Nichol}, R.~C. and {Parkinson}, H.~R. and {Phillipps}, S. and {Pimbblet}, K. and {Popescu}, C.~C. and {Prescott}, M. and {Roseboom}, I.~G. and {Sadler}, E.~M. and {Sansom}, A.~E. and {Sharp}, R.~G. and {Smith}, D.~J.~B. and {Taylor}, E. and {Thomas}, D. and {Tuffs}, R.~J. and {Wijesinghe}, D. and {Dunne}, L. and {Frenk}, C.~S. and {Jarvis}, M.~J. and {Madore}, B.~F. and {Meyer}, M.~J. and {Seibert}, M. and {Staveley-Smith}, L. and {Sutherland}, W.~J. and {Warren}, S.~J.},
        title = "{Galaxy and Mass Assembly (GAMA): survey diagnostics and core data release}",
      journal = {MNRAS},
         year = 2011,
        month = may,
       volume = {413},
       number = {2},
        pages = {971-995},
          doi = {10.1111/j.1365-2966.2010.18188.x},
archivePrefix = {arXiv},
       eprint = {1009.0614},
 primaryClass = {astro-ph.CO},
       adsurl = {https://ui.adsabs.harvard.edu/abs/2011MNRAS.413..971D}
}

@ARTICLE{2010AJ....140.1868W,
       author = {{Wright}, Edward L. and {Eisenhardt}, Peter R.~M. and {Mainzer}, Amy K. and {Ressler}, Michael E. and {Cutri}, Roc M. and {Jarrett}, Thomas and {Kirkpatrick}, J. Davy and {Padgett}, Deborah and {McMillan}, Robert S. and {Skrutskie}, Michael and {Stanford}, S.~A. and {Cohen}, Martin and {Walker}, Russell G. and {Mather}, John C. and {Leisawitz}, David and {Gautier}, III, Thomas N. and {McLean}, Ian and {Benford}, Dominic and {Lonsdale}, Carol J. and {Blain}, Andrew and {Mendez}, Bryan and {Irace}, William R. and {Duval}, Valerie and {Liu}, Fengchuan and {Royer}, Don and {Heinrichsen}, Ingolf and {Howard}, Joan and {Shannon}, Mark and {Kendall}, Martha and {Walsh}, Amy L. and {Larsen}, Mark and {Cardon}, Joel G. and {Schick}, Scott and {Schwalm}, Mark and {Abid}, Mohamed and {Fabinsky}, Beth and {Naes}, Larry and {Tsai}, Chao-Wei},
        title = "{The Wide-field Infrared Survey Explorer (WISE): Mission Description and Initial On-orbit Performance}",
      journal = {AJ},
         year = 2010,
        month = dec,
       volume = {140},
       number = {6},
        pages = {1868-1881},
          doi = {10.1088/0004-6256/140/6/1868},
archivePrefix = {arXiv},
       eprint = {1008.0031},
 primaryClass = {astro-ph.IM},
       adsurl = {https://ui.adsabs.harvard.edu/abs/2010AJ....140.1868W}
}

@ARTICLE{2020ApJ...903...91Y,
       author = {{Yao}, H.~F.~M. and {Jarrett}, T.~H. and {Cluver}, M.~E. and {Marchetti}, L. and {Taylor}, Edward N. and {Santos}, M.~G. and {Owers}, Matt S. and {Lopez-Sanchez}, Angel R. and {Gordon}, Y.~A. and {Brown}, M.~J.~I. and et al.},
        title = "{Galaxy and Mass Assembly (GAMA): A WISE Study of the Activity of Emission-line Systems in G23}",
      journal = {\apj},
         year = 2020,
        month = nov,
       volume = {903},
       number = {2},
          eid = {91},
        pages = {91},
          doi = {10.3847/1538-4357/abba1a},
archivePrefix = {arXiv},
       eprint = {2009.05981},
 primaryClass = {astro-ph.GA},
       adsurl = {https://ui.adsabs.harvard.edu/abs/2020ApJ...903...91Y}
}

@ARTICLE{2014ApJ...782...90C,
       author = {{Cluver}, M.~E. and {Jarrett}, T.~H. and {Hopkins}, A.~M. and {Driver}, S.~P. and {Liske}, J. and {Gunawardhana}, M.~L.~P. and {Taylor}, E.~N. and {Robotham}, A.~S.~G. and {Alpaslan}, M. and {Baldry}, I. and {Brown}, M.~J.~I. and {Peacock}, J.~A. and {Popescu}, C.~C. and {Tuffs}, R.~J. and {Bauer}, A.~E. and {Bland-Hawthorn}, J. and {Colless}, M. and {Holwerda}, B.~W. and {Lara-L{\'o}pez}, M.~A. and {Leschinski}, K. and {L{\'o}pez-S{\'a}nchez}, A.~R. and {Norberg}, P. and {Owers}, M.~S. and {Wang}, L. and {Wilkins}, S.~M.},
        title = "{Galaxy and Mass Assembly (GAMA): Mid-infrared Properties and Empirical Relations from WISE}",
      journal = {ApJ},
         year = 2014,
        month = feb,
       volume = {782},
       number = {2},
          eid = {90},
        pages = {90},
          doi = {10.1088/0004-637X/782/2/90},
archivePrefix = {arXiv},
       eprint = {1401.0837},
 primaryClass = {astro-ph.GA},
       adsurl = {https://ui.adsabs.harvard.edu/abs/2014ApJ...782...90C}
}

@ARTICLE{2019PASP..131a8002B,
       author = {{Bellm}, Eric C. and {Kulkarni}, Shrinivas R. and {Graham}, Matthew J. and {Dekany}, Richard and {Smith}, Roger M. and {Riddle}, Reed and {Masci}, Frank J. and {Helou}, George and {Prince}, Thomas A. and {Adams}, Scott M. and {Barbarino}, C. and {Barlow}, Tom and {Bauer}, James and {Beck}, Ron and {Belicki}, Justin and {Biswas}, Rahul and {Blagorodnova}, Nadejda and {Bodewits}, Dennis and {Bolin}, Bryce and {Brinnel}, Valery and {Brooke}, Tim and {Bue}, Brian and {Bulla}, Mattia and {Burruss}, Rick and {Cenko}, S. Bradley and {Chang}, Chan-Kao and {Connolly}, Andrew and {Coughlin}, Michael and {Cromer}, John and {Cunningham}, Virginia and {De}, Kishalay and {Delacroix}, Alex and {Desai}, Vandana and {Duev}, Dmitry A. and {Eadie}, Gwendolyn and {Farnham}, Tony L. and {Feeney}, Michael and {Feindt}, Ulrich and {Flynn}, David and {Franckowiak}, Anna and {Frederick}, S. and {Fremling}, C. and {Gal-Yam}, Avishay and {Gezari}, Suvi and {Giomi}, Matteo and {Goldstein}, Daniel A. and {Golkhou}, V. Zach and {Goobar}, Ariel and {Groom}, Steven and {Hacopians}, Eugean and {Hale}, David and {Henning}, John and {Ho}, Anna Y.~Q. and {Hover}, David and {Howell}, Justin and {Hung}, Tiara and {Huppenkothen}, Daniela and {Imel}, David and {Ip}, Wing-Huen and {Ivezi{\'c}}, {\v{Z}}eljko and {Jackson}, Edward and {Jones}, Lynne and {Juric}, Mario and {Kasliwal}, Mansi M. and {Kaspi}, S. and {Kaye}, Stephen and {Kelley}, Michael S.~P. and {Kowalski}, Marek and {Kramer}, Emily and {Kupfer}, Thomas and {Landry}, Walter and {Laher}, Russ R. and {Lee}, Chien-De and {Lin}, Hsing Wen and {Lin}, Zhong-Yi and {Lunnan}, Ragnhild and {Giomi}, Matteo and {Mahabal}, Ashish and {Mao}, Peter and {Miller}, Adam A. and {Monkewitz}, Serge and {Murphy}, Patrick and {Ngeow}, Chow-Choong and {Nordin}, Jakob and {Nugent}, Peter and {Ofek}, Eran and {Patterson}, Maria T. and {Penprase}, Bryan and {Porter}, Michael and {Rauch}, Ludwig and {Rebbapragada}, Umaa and {Reiley}, Dan and {Rigault}, Mickael and {Rodriguez}, Hector and {van Roestel}, Jan and {Rusholme}, Ben and {van Santen}, Jakob and {Schulze}, S. and {Shupe}, David L. and {Singer}, Leo P. and {Soumagnac}, Maayane T. and {Stein}, Robert and {Surace}, Jason and {Sollerman}, Jesper and {Szkody}, Paula and {Taddia}, F. and {Terek}, Scott and {Van Sistine}, Angela and {van Velzen}, Sjoert and {Vestrand}, W. Thomas and {Walters}, Richard and {Ward}, Charlotte and {Ye}, Quan-Zhi and {Yu}, Po-Chieh and {Yan}, Lin and {Zolkower}, Jeffry},
        title = "{The Zwicky Transient Facility: System Overview, Performance, and First Results}",
      journal = {PASP},
         year = 2019,
        month = jan,
       volume = {131},
       number = {995},
        pages = {018002},
          doi = {10.1088/1538-3873/aaecbe},
archivePrefix = {arXiv},
       eprint = {1902.01932},
 primaryClass = {astro-ph.IM},
       adsurl = {https://ui.adsabs.harvard.edu/abs/2019PASP..131a8002B}
}

@ARTICLE{2009ApJ...696..870D,
       author = {{Drake}, A.~J. and {Djorgovski}, S.~G. and {Mahabal}, A. and {Beshore}, E. and {Larson}, S. and {Graham}, M.~J. and {Williams}, R. and {Christensen}, E. and {Catelan}, M. and {Boattini}, A. and {Gibbs}, A. and {Hill}, R. and {Kowalski}, R.},
        title = "{First Results from the Catalina Real-Time Transient Survey}",
      journal = {ApJ},
         year = 2009,
        month = may,
       volume = {696},
       number = {1},
        pages = {870-884},
          doi = {10.1088/0004-637X/696/1/870},
archivePrefix = {arXiv},
       eprint = {0809.1394},
 primaryClass = {astro-ph},
       adsurl = {https://ui.adsabs.harvard.edu/abs/2009ApJ...696..870D}
}

@ARTICLE{2011ApJ...731...53M,
       author = {{Mainzer}, A. and {Bauer}, J. and {Grav}, T. and {Masiero}, J. and {Cutri}, R.~M. and {Dailey}, J. and {Eisenhardt}, P. and {McMillan}, R.~S. and {Wright}, E. and {Walker}, R. and {Jedicke}, R. and {Spahr}, T. and {Tholen}, D. and {Alles}, R. and {Beck}, R. and {Brandenburg}, H. and {Conrow}, T. and {Evans}, T. and {Fowler}, J. and {Jarrett}, T. and {Marsh}, K. and {Masci}, F. and {McCallon}, H. and {Wheelock}, S. and {Wittman}, M. and {Wyatt}, P. and {DeBaun}, E. and {Elliott}, G. and {Elsbury}, D. and {Gautier}, IV, T. and {Gomillion}, S. and {Leisawitz}, D. and {Maleszewski}, C. and {Micheli}, M. and {Wilkins}, A.},
        title = "{Preliminary Results from NEOWISE: An Enhancement to the Wide-field Infrared Survey Explorer for Solar System Science}",
      journal = {ApJ},
         year = 2011,
        month = apr,
       volume = {731},
       number = {1},
          eid = {53},
        pages = {53},
          doi = {10.1088/0004-637X/731/1/53},
archivePrefix = {arXiv},
       eprint = {1102.1996},
 primaryClass = {astro-ph.EP},
       adsurl = {https://ui.adsabs.harvard.edu/abs/2011ApJ...731...53M}
}

@ARTICLE{2024ApJ...966...85Z,
       author = {{Zeltyn}, Grisha and {Trakhtenbrot}, Benny and {Eracleous}, Michael and {Yang}, Qian and {Green}, Paul and {Anderson}, Scott F. and {LaMassa}, Stephanie and {Runnoe}, Jessie and {Assef}, Roberto J. and {Bauer}, Franz E. and {Brandt}, W.~N. and {Davis}, Megan C. and {Frederick}, Sara E. and {Fries}, Logan B. and {Graham}, Matthew J. and {Grogin}, Norman A. and {Guolo}, Muryel and {Hern{\'a}ndez-Garc{\'\i}a}, Lorena and {Koekemoer}, Anton M. and {Krumpe}, Mirko and {Liu}, Xin and {Mart{\'\i}nez-Aldama}, Mary Loli and {Ricci}, Claudio and {Schneider}, Donald P. and {Shen}, Yue and {{\'S}niegowska}, Marzena and {Temple}, Matthew J. and {Trump}, Jonathan R. and {Xue}, Yongquan and {Brownstein}, Joel R. and {Dwelly}, Tom and {Morrison}, Sean and {Bizyaev}, Dmitry and {Pan}, Kaike and {Kollmeier}, Juna A.},
        title = "{Exploring Changing-look Active Galactic Nuclei with the Sloan Digital Sky Survey V: First Year Results}",
      journal = {ApJ},
         year = 2024,
        month = may,
       volume = {966},
       number = {1},
          eid = {85},
        pages = {85},
          doi = {10.3847/1538-4357/ad2f30},
archivePrefix = {arXiv},
       eprint = {2401.01933},
 primaryClass = {astro-ph.GA},
       adsurl = {https://ui.adsabs.harvard.edu/abs/2024ApJ...966...85Z}
}

@ARTICLE{2013AJ....146...32S,
       author = {{Smee}, Stephen A. and {Gunn}, James E. and {Uomoto}, Alan and {Roe}, Natalie and {Schlegel}, David and {Rockosi}, Constance M. and {Carr}, Michael A. and {Leger}, French and {Dawson}, Kyle S. and {Olmstead}, Matthew D. and {Brinkmann}, Jon and {Owen}, Russell and {Barkhouser}, Robert H. and {Honscheid}, Klaus and {Harding}, Paul and {Long}, Dan and {Lupton}, Robert H. and {Loomis}, Craig and {Anderson}, Lauren and {Annis}, James and {Bernardi}, Mariangela and {Bhardwaj}, Vaishali and {Bizyaev}, Dmitry and {Bolton}, Adam S. and {Brewington}, Howard and {Briggs}, John W. and {Burles}, Scott and {Burns}, James G. and {Castander}, Francisco Javier and {Connolly}, Andrew and {Davenport}, James R.~A. and {Ebelke}, Garrett and {Epps}, Harland and {Feldman}, Paul D. and {Friedman}, Scott D. and {Frieman}, Joshua and {Heckman}, Timothy and {Hull}, Charles L. and {Knapp}, Gillian R. and {Lawrence}, David M. and {Loveday}, Jon and {Mannery}, Edward J. and {Malanushenko}, Elena and {Malanushenko}, Viktor and {Merrelli}, Aronne James and {Muna}, Demitri and {Newman}, Peter R. and {Nichol}, Robert C. and {Oravetz}, Daniel and {Pan}, Kaike and {Pope}, Adrian C. and {Ricketts}, Paul G. and {Shelden}, Alaina and {Sandford}, Dale and {Siegmund}, Walter and {Simmons}, Audrey and {Smith}, D. Shane and {Snedden}, Stephanie and {Schneider}, Donald P. and {SubbaRao}, Mark and {Tremonti}, Christy and {Waddell}, Patrick and {York}, Donald G.},
        title = "{The Multi-object, Fiber-fed Spectrographs for the Sloan Digital Sky Survey and the Baryon Oscillation Spectroscopic Survey}",
      journal = {AJ},
         year = 2013,
        month = aug,
       volume = {146},
       number = {2},
          eid = {32},
        pages = {32},
          doi = {10.1088/0004-6256/146/2/32},
archivePrefix = {arXiv},
       eprint = {1208.2233},
 primaryClass = {astro-ph.IM},
       adsurl = {https://ui.adsabs.harvard.edu/abs/2013AJ....146...32S}
}

@ARTICLE{2025arXiv250707093S,
       author = {{SDSS Collaboration} and {Adamane Pallathadka}, Gautham and {Aghakhanloo}, Mojgan and {Aird}, James and {Almeida}, Andr{\'e}s and {Amrita}, Singh and {Anders}, Friedrich and {Anderson}, Scott F. and {Arseneau}, Stefan and {Gonz{\'a}lez Avila}, Consuelo and {Aviram}, Shir and {Aydar}, Catarina and {Badenes}, Carles and {Barrera-Ballesteros}, Jorge K. and {Bauer}, Franz E. and {Behmard}, Aida and {Berg}, Michelle and {Besser}, F. and {Moni Bidin}, Christian and {Bizyaev}, Dmitry and {Blanc}, Guillermo and {Blanton}, Michael R. and {Bovy}, Jo and {Brandt}, William Nielsen and {Brownstein}, Joel R. and {Buchner}, Johannes and {Bulbul}, Esra and {Burchett}, Joseph N. and {Carigi}, Leticia and {Carlberg}, Joleen K. and {Casey}, Andrew R. and {Chakraborty}, Priyanka and {Chanam{\'e}}, Julio and {Chandra}, Vedant and {Chiappini}, Cristina and {Chilingarian}, Igor and {Comparat}, Johan and {Covey}, Kevin and {Crumpler}, Nicole and {Cunha}, Katia and {D'Onghia}, Elena and {Dai}, Xinyu and {Darling}, Jeremy and {Davis}, Megan and {De Lee}, Nathan and {Deacon}, Niall and {M{\'e}ndez Delgado}, Jos{\'e} Eduardo and {Demasi}, Sebastian and {Demianenko}, Mariia and {Demke}, Delvin and {Donor}, John and {Drory}, Niv and {Villa Durango}, Monica Alejandra and {Dwelly}, Tom and {Egorov}, Oleg and {Egorova}, Evgeniya and {El-Badry}, Kareem and {Eracleous}, Mike and {Fan}, Xiaohui and {Farr}, Emily and {Finkbeiner}, Douglas P. and {Fries}, Logan and {Frinchaboy}, Peter and {Gentile Fusillo}, Nicola Pietro and {Serrano F{\'e}lix}, Luis Daniel and {Gaensicke}, Boris and {Galligan}, Emma and {Garc{\'\i}a}, Pablo and {Gelfand}, Joseph and {Grabowski}, Katie and {Grebel}, Eva and {Green}, Paul J and {Greve}, Hannah and {Grier}, Catherine and {Griffith}, Emily and {Guetzoyan}, Paloma and {Gupta}, Pramod and {Hackshaw}, Zoe and {Hall}, Patrick B. and {Hawkins}, Keith and {Heged{\H{u}}s}, Viola and {Hekker}, Saskia and {Herbst}, T.~M. and {Hermes}, J.~J. and {Hern{\'a}ndez-Garc{\'\i}a}, Lorena and {Hiremath}, Pranavi and {Hogg}, David W and {Holtzman}, Jon and {Horne}, Keith and {Horta}, Danny and {Huang}, Yang and {Hutchinson}, Brian and {H{\"a}berle}, Maximilian and {Ibarra-Medel}, Hector Javier and {Ji}, Alexander P. and {Jofre}, Paula and {Johnson}, James W. and {Johnson}, Jennifer and {Johnston}, Evelyn J. and {Kaldor}, Mary and {Katkov}, Ivan and {Khalatyan}, Arman and {Khoperskov}, Sergey and {Klessen}, Ralf and {Kluge}, Matthias and {Koekemoer}, Anton M. and {Kollmeier}, Juna A. and {Kounkel}, Marina and {Kreckel}, Kathryn and {Krishnarao}, Dhanesh and {Krumpe}, Mirko and {Lacerna}, Ivan and {Laporte}, Chervin and {Lepine}, Sebastien and {Li}, Jing and {Liang}, Fu-Heng and {Limberg}, Guilherme and {Liu}, Xin and {Loebman}, Sarah and {Long}, Knox and {Lu}, Yuxi and {Lucey}, Madeline and {Lugo-Aranda}, Alejandra Z. and {Mart{\'\i}nez Martinez-Aldama}, Mary Loli and {McKinnon}, Kevin and {Medan}, Ilija and {Merloni}, Andrea and {Morrison}, Sean and {Myers}, Natalie and {M{\'e}sz{\'a}ros}, Szabolcs and {M{\"u}ller-Horn}, Johanna and {Nepal}, Samir and {Ness}, Melissa and {Nidever}, David and {Nitschelm}, Christian and {Oravetz}, Audrey and {Otto}, Jonah and {Pan}, Kaike and {P{\'e}rez Paolino}, Facundo and {Negrete Pe{\~n}aloza}, Castalia Alenka and {Pinsonneault}, Marc and {Taghizadeh Popp}, Manuchehr and {Price-Whelan}, Adrian and {Pulatova}, Nadiia and {Queiroz}, Anna Barbara and {Raddick}, Jordan and {Rankine}, Amy and {Rix}, Hans-Walter and {Rom{\'a}n-Z{\'u}{\~n}iga}, Carlos and {Fern{\'a}ndez Rosso}, Daniela and {Runnoe}, Jessie and {Mahmud Saad}, Serat and {Salvato}, Mara and {Sanchez}, Sebastian F. and {Sattler}, Natascha and {Saydjari}, Andrew and {Sayres}, Conor and {Schlaufman}, Kevin and {Schneider}, Donald P. and {Schwope}, Axel and {Seaton}, Lucas M. and {Seeburger}, Rhys and {Serna}, Javier and {Sharma}, Sanjib and {Shen}, Yue and {Sinha}, Amaya and {Sizemore}, Brian and {Sniegowska}, Marzena and {Song}, Yingyi and {Souto}, Diogo and {Stassun}, Keivan and {Steinmetz}, Matthias and {Stone}, Zachary and {Stone-Martinez}, Alexander and {Stringfellow}, Guy S. and {Mata S{\'a}nchez}, Aurora and {S{\'a}nchez-Gallego}, Jos{\'e} and {Tan}, Jonathan and {Tayar}, Jamie and {Thai}, Riley and {Thakar}, Ani and {Thibodeaux}, Pierre and {Ting}, Yuan-Sen and {Tkachenko}, Andrew and {Trakhtenbrot}, Benny and {Fernandez Trincado}, Jose G. and {Troup}, Nicholas and {Trump}, Jonathan R. and {Ulloa}, Natalie and {Van der Marel}, Roeland P. and {Vera}, Pablo and {Villanova}, Sandro and {Villase{\~n}or}, Jaime and {Wang}, Ji and {Way}, Zachary and {Weijmans}, Anne-Marie and {Wheeler}, Adam and {Wilson}, John C. and {Wofford}, Aida and {Wong}, Tony},
        title = "{The Nineteenth Data Release of the Sloan Digital Sky Survey}",
      journal = {arXiv},
         year = 2025,
        month = jul,
          eid = {arXiv:2507.07093},
        pages = {arXiv:2507.07093},
          doi = {10.48550/arXiv.2507.07093},
archivePrefix = {arXiv},
       eprint = {2507.07093},
 primaryClass = {astro-ph.GA},
       adsurl = {https://ui.adsabs.harvard.edu/abs/2025arXiv250707093S}
}

@ARTICLE{2022PASJ...74..247A,
       author = {{Aihara}, Hiroaki and {AlSayyad}, Yusra and {Ando}, Makoto and {Armstrong}, Robert and {Bosch}, James and {Egami}, Eiichi and {Furusawa}, Hisanori and {Furusawa}, Junko and {Harasawa}, Sumiko and {Harikane}, Yuichi and {Hsieh}, Bau-Ching and {Ikeda}, Hiroyuki and {Ito}, Kei and {Iwata}, Ikuru and {Kodama}, Tadayuki and {Koike}, Michitaro and {Kokubo}, Mitsuru and {Komiyama}, Yutaka and {Li}, Xiangchong and {Liang}, Yongming and {Lin}, Yen-Ting and {Lupton}, Robert H. and {Lust}, Nate B. and {MacArthur}, Lauren A. and {Mawatari}, Ken and {Mineo}, Sogo and {Miyatake}, Hironao and {Miyazaki}, Satoshi and {More}, Surhud and {Morishima}, Takahiro and {Murayama}, Hitoshi and {Nakajima}, Kimihiko and {Nakata}, Fumiaki and {Nishizawa}, Atsushi J. and {Oguri}, Masamune and {Okabe}, Nobuhiro and {Okura}, Yuki and {Ono}, Yoshiaki and {Osato}, Ken and {Ouchi}, Masami and {Pan}, Yen-Chen and {Plazas Malag{\'o}n}, Andr{\'e}s A. and {Price}, Paul A. and {Reed}, Sophie L. and {Rykoff}, Eli S. and {Shibuya}, Takatoshi and {Simunovic}, Mirko and {Strauss}, Michael A. and {Sugimori}, Kanako and {Suto}, Yasushi and {Suzuki}, Nao and {Takada}, Masahiro and {Takagi}, Yuhei and {Takata}, Tadafumi and {Takita}, Satoshi and {Tanaka}, Masayuki and {Tang}, Shenli and {Taranu}, Dan S. and {Terai}, Tsuyoshi and {Toba}, Yoshiki and {Turner}, Edwin L. and {Uchiyama}, Hisakazu and {Vijarnwannaluk}, Bovornpratch and {Waters}, Christopher Z. and {Yamada}, Yoshihiko and {Yamamoto}, Naoaki and {Yamashita}, Takuji},
        title = "{Third data release of the Hyper Suprime-Cam Subaru Strategic Program}",
      journal = {PASJ},
         year = 2022,
        month = apr,
       volume = {74},
       number = {2},
        pages = {247-272},
          doi = {10.1093/pasj/psab122},
archivePrefix = {arXiv},
       eprint = {2108.13045},
 primaryClass = {astro-ph.IM},
       adsurl = {https://ui.adsabs.harvard.edu/abs/2022PASJ...74..247A}
}

@ARTICLE{2015MNRAS.447.2857B,
       author = {{Bryant}, J.~J. and {Owers}, M.~S. and {Robotham}, A.~S.~G. and {Croom}, S.~M. and {Driver}, S.~P. and {Drinkwater}, M.~J. and {Lorente}, N.~P.~F. and {Cortese}, L. and {Scott}, N. and {Colless}, M. and {Schaefer}, A. and {Taylor}, E.~N. and {Konstantopoulos}, I.~S. and {Allen}, J.~T. and {Baldry}, I. and {Barnes}, L. and {Bauer}, A.~E. and {Bland-Hawthorn}, J. and {Bloom}, J.~V. and {Brooks}, A.~M. and {Brough}, S. and {Cecil}, G. and {Couch}, W. and {Croton}, D. and {Davies}, R. and {Ellis}, S. and {Fogarty}, L.~M.~R. and {Foster}, C. and {Glazebrook}, K. and {Goodwin}, M. and {Green}, A. and {Gunawardhana}, M.~L. and {Hampton}, E. and {Ho}, I.-T. and {Hopkins}, A.~M. and {Kewley}, L. and {Lawrence}, J.~S. and {Leon-Saval}, S.~G. and {Leslie}, S. and {McElroy}, R. and {Lewis}, G. and {Liske}, J. and {L{\'o}pez-S{\'a}nchez}, {\'A}. R. and {Mahajan}, S. and {Medling}, A.~M. and {Metcalfe}, N. and {Meyer}, M. and {Mould}, J. and {Obreschkow}, D. and {O'Toole}, S. and {Pracy}, M. and {Richards}, S.~N. and {Shanks}, T. and {Sharp}, R. and {Sweet}, S.~M. and {Thomas}, A.~D. and {Tonini}, C. and {Walcher}, C.~J.},
        title = "{The SAMI Galaxy Survey: instrument specification and target selection}",
      journal = {MNRAS},
         year = 2015,
        month = mar,
       volume = {447},
       number = {3},
        pages = {2857-2879},
          doi = {10.1093/mnras/stu2635},
archivePrefix = {arXiv},
       eprint = {1407.7335},
 primaryClass = {astro-ph.GA},
       adsurl = {https://ui.adsabs.harvard.edu/abs/2015MNRAS.447.2857B}
}

@ARTICLE{2011MNRAS.415.2173B,
       author = {{Bryant}, J.~J. and {O'Byrne}, J.~W. and {Bland-Hawthorn}, J. and {Leon-Saval}, S.~G.},
        title = "{Characterization of hexabundles: initial results}",
      journal = {MNRAS},
         year = 2011,
        month = aug,
       volume = {415},
       number = {3},
        pages = {2173-2181},
          doi = {10.1111/j.1365-2966.2011.18841.x},
archivePrefix = {arXiv},
       eprint = {1104.0985},
 primaryClass = {astro-ph.IM},
       adsurl = {https://ui.adsabs.harvard.edu/abs/2011MNRAS.415.2173B}
}

@ARTICLE{2012MNRAS.421..872C,
       author = {{Croom}, Scott M. and {Lawrence}, Jon S. and {Bland-Hawthorn}, Joss and {Bryant}, Julia J. and {Fogarty}, Lisa and {Richards}, Samuel and {Goodwin}, Michael and {Farrell}, Tony and {Miziarski}, Stan and {Heald}, Ron and {Jones}, D. Heath and {Lee}, Steve and {Colless}, Matthew and {Brough}, Sarah and {Hopkins}, Andrew M. and {Bauer}, Amanda E. and {Birchall}, Michael N. and {Ellis}, Simon and {Horton}, Anthony and {Leon-Saval}, Sergio and {Lewis}, Geraint and {L{\'o}pez-S{\'a}nchez}, {\'A}. R. and {Min}, Seong-Sik and {Trinh}, Christopher and {Trowland}, Holly},
        title = "{The Sydney-AAO Multi-object Integral field spectrograph}",
      journal = {MNRAS},
         year = 2012,
        month = mar,
       volume = {421},
       number = {1},
        pages = {872-893},
          doi = {10.1111/j.1365-2966.2011.20365.x},
archivePrefix = {arXiv},
       eprint = {1112.3367},
 primaryClass = {astro-ph.CO},
       adsurl = {https://ui.adsabs.harvard.edu/abs/2012MNRAS.421..872C}
}

@ARTICLE{2009ApJS..182..543A,
       author = {{Abazajian}, Kevork N. and {Adelman-McCarthy}, Jennifer K. and {Ag{\"u}eros}, Marcel A. and {Allam}, Sahar S. and {Allende Prieto}, Carlos and {An}, Deokkeun and {Anderson}, Kurt S.~J. and {Anderson}, Scott F. and {Annis}, James and {Bahcall}, Neta A. and {Bailer-Jones}, C.~A.~L. and {Barentine}, J.~C. and {Bassett}, Bruce A. and {Becker}, Andrew C. and {Beers}, Timothy C. and {Bell}, Eric F. and {Belokurov}, Vasily and {Berlind}, Andreas A. and {Berman}, Eileen F. and {Bernardi}, Mariangela and {Bickerton}, Steven J. and {Bizyaev}, Dmitry and {Blakeslee}, John P. and {Blanton}, Michael R. and {Bochanski}, John J. and {Boroski}, William N. and {Brewington}, Howard J. and {Brinchmann}, Jarle and {Brinkmann}, J. and {Brunner}, Robert J. and {Budav{\'a}ri}, Tam{\'a}s and {Carey}, Larry N. and {Carliles}, Samuel and {Carr}, Michael A. and {Castander}, Francisco J. and {Cinabro}, David and {Connolly}, A.~J. and {Csabai}, Istv{\'a}n and {Cunha}, Carlos E. and {Czarapata}, Paul C. and {Davenport}, James R.~A. and {de Haas}, Ernst and {Dilday}, Ben and {Doi}, Mamoru and {Eisenstein}, Daniel J. and {Evans}, Michael L. and {Evans}, N.~W. and {Fan}, Xiaohui and {Friedman}, Scott D. and {Frieman}, Joshua A. and {Fukugita}, Masataka and {G{\"a}nsicke}, Boris T. and {Gates}, Evalyn and {Gillespie}, Bruce and {Gilmore}, G. and {Gonzalez}, Belinda and {Gonzalez}, Carlos F. and {Grebel}, Eva K. and {Gunn}, James E. and {Gy{\"o}ry}, Zsuzsanna and {Hall}, Patrick B. and {Harding}, Paul and {Harris}, Frederick H. and {Harvanek}, Michael and {Hawley}, Suzanne L. and {Hayes}, Jeffrey J.~E. and {Heckman}, Timothy M. and {Hendry}, John S. and {Hennessy}, Gregory S. and {Hindsley}, Robert B. and {Hoblitt}, J. and {Hogan}, Craig J. and {Hogg}, David W. and {Holtzman}, Jon A. and {Hyde}, Joseph B. and {Ichikawa}, Shin-ichi and {Ichikawa}, Takashi and {Im}, Myungshin and {Ivezi{\'c}}, {\v{Z}}eljko and {Jester}, Sebastian and {Jiang}, Linhua and {Johnson}, Jennifer A. and {Jorgensen}, Anders M. and {Juri{\'c}}, Mario and {Kent}, Stephen M. and {Kessler}, R. and {Kleinman}, S.~J. and {Knapp}, G.~R. and {Konishi}, Kohki and {Kron}, Richard G. and {Krzesinski}, Jurek and {Kuropatkin}, Nikolay and {Lampeitl}, Hubert and {Lebedeva}, Svetlana and {Lee}, Myung Gyoon and {Lee}, Young Sun and {French Leger}, R. and {L{\'e}pine}, S{\'e}bastien and {Li}, Nolan and {Lima}, Marcos and {Lin}, Huan and {Long}, Daniel C. and {Loomis}, Craig P. and {Loveday}, Jon and {Lupton}, Robert H. and {Magnier}, Eugene and {Malanushenko}, Olena and {Malanushenko}, Viktor and {Mandelbaum}, Rachel and {Margon}, Bruce and {Marriner}, John P. and {Mart{\'\i}nez-Delgado}, David and {Matsubara}, Takahiko and {McGehee}, Peregrine M. and {McKay}, Timothy A. and {Meiksin}, Avery and {Morrison}, Heather L. and {Mullally}, Fergal and {Munn}, Jeffrey A. and {Murphy}, Tara and {Nash}, Thomas and {Nebot}, Ada and {Neilsen}, Jr., Eric H. and {Newberg}, Heidi Jo and {Newman}, Peter R. and {Nichol}, Robert C. and {Nicinski}, Tom and {Nieto-Santisteban}, Maria and {Nitta}, Atsuko and {Okamura}, Sadanori and {Oravetz}, Daniel J. and {Ostriker}, Jeremiah P. and {Owen}, Russell and {Padmanabhan}, Nikhil and {Pan}, Kaike and {Park}, Changbom and {Pauls}, George and {Peoples}, Jr., John and {Percival}, Will J. and {Pier}, Jeffrey R. and {Pope}, Adrian C. and {Pourbaix}, Dimitri and {Price}, Paul A. and {Purger}, Norbert and {Quinn}, Thomas and {Raddick}, M. Jordan and {Re Fiorentin}, Paola and {Richards}, Gordon T. and {Richmond}, Michael W. and {Riess}, Adam G. and {Rix}, Hans-Walter and {Rockosi}, Constance M. and {Sako}, Masao and {Schlegel}, David J. and {Schneider}, Donald P. and {Scholz}, Ralf-Dieter and {Schreiber}, Matthias R. and {Schwope}, Axel D. and {Seljak}, Uro{\v{s}} and {Sesar}, Branimir and {Sheldon}, Erin and {Shimasaku}, Kazu and {Sibley}, Valena C. and {Simmons}, A.~E. and {Sivarani}, Thirupathi and {Allyn Smith}, J. and {Smith}, Martin C. and {Smol{\v{c}}i{\'c}}, Vernesa and {Snedden}, Stephanie A. and {Stebbins}, Albert and {Steinmetz}, Matthias and {Stoughton}, Chris and {Strauss}, Michael A. and {SubbaRao}, Mark and {Suto}, Yasushi and {Szalay}, Alexander S. and {Szapudi}, Istv{\'a}n and {Szkody}, Paula and {Tanaka}, Masayuki and {Tegmark}, Max and {Teodoro}, Luis F.~A. and {Thakar}, Aniruddha R. and {Tremonti}, Christy A. and {Tucker}, Douglas L. and {Uomoto}, Alan and {Vanden Berk}, Daniel E. and {Vandenberg}, Jan and {Vidrih}, S. and {Vogeley}, Michael S. and {Voges}, Wolfgang and {Vogt}, Nicole P. and {Wadadekar}, Yogesh and {Watters}, Shannon and {Weinberg}, David H. and {West}, Andrew A. and {White}, Simon D.~M. and {Wilhite}, Brian C. and {Wonders}, Alainna C. and {Yanny}, Brian and {Yocum}, D.~R.},
        title = "{The Seventh Data Release of the Sloan Digital Sky Survey}",
      journal = {ApJS},
         year = 2009,
        month = jun,
       volume = {182},
       number = {2},
        pages = {543-558},
          doi = {10.1088/0067-0049/182/2/543},
archivePrefix = {arXiv},
       eprint = {0812.0649},
 primaryClass = {astro-ph},
       adsurl = {https://ui.adsabs.harvard.edu/abs/2009ApJS..182..543A}
}

@ARTICLE{1984PASP...96..393G,
       author = {{Gaskell}, C.~M. and {Ferland}, G.~J.},
        title = "{Theoretical hydrogen-line ratios for the narrow-line regions of active galactic nuclei}",
      journal = {PASP},
         year = 1984,
        month = jun,
       volume = {96},
        pages = {393-397},
          doi = {10.1086/131352},
       adsurl = {https://ui.adsabs.harvard.edu/abs/1984PASP...96..393G}
}

@ARTICLE{2008MNRAS.383..581D,
       author = {{Dong}, Xiaobo and {Wang}, Tinggui and {Wang}, Jianguo and {Yuan}, Weimin and {Zhou}, Hongyan and {Dai}, Haifeng and {Zhang}, Kai},
        title = "{Broad-line Balmer decrements in blue active galactic nuclei}",
      journal = {MNRAS},
         year = 2008,
        month = jan,
       volume = {383},
       number = {2},
        pages = {581-592},
          doi = {10.1111/j.1365-2966.2007.12560.x},
archivePrefix = {arXiv},
       eprint = {0710.1458},
 primaryClass = {astro-ph},
       adsurl = {https://ui.adsabs.harvard.edu/abs/2008MNRAS.383..581D}
}

\end{document}